\documentclass[aip, pof, preprint]{revtex4-1}
\usepackage{graphicx}
\usepackage{dcolumn}
\usepackage{bm}
\usepackage{amsmath}
\usepackage{mathtools}
\usepackage{epstopdf}
\usepackage{color}
\usepackage{soul}
\begin{document}
%
%
\title{DROPLET SPREADING ON ROUGH SURFACES: TACKLING THE CONTACT LINE BOUNDARY CONDITION}
\author{N. T. Chamakos}
\author{M. E. Kavousanakis}
\author{A. G. Boudouvis}
\author{A. G. Papathanasiou}
\email{pathan@chemeng.ntua.gr}
\affiliation{School of Chemical Engineering, National Technical University of Athens, 15780, Greece}
\date{\today}
%
%
\begin{abstract}
The complicated dynamics of the contact line of a moving droplet on a solid substrate often hamper the efficient modeling of microfluidic systems. In particular, the selection of the effective boundary conditions, specifying the contact line motion, is a controversial issue since the microscopic physics that gives rise to this displacement is still unknown. Here, a sharp interface, continuum-level, novel modeling approach, accounting for liquid/solid micro-scale interactions assembled in a disjoining pressure term, is presented. By following a unified conception (the model applies both to the liquid/solid and the liquid/ambient interfaces), the friction forces at the contact line, as well as the dynamic contact angle are derived implicitly as a result of the disjoining pressure and viscous effects interplay in the vicinity of the substrate's intrinsic roughness. Previous hydrodynamic model limitations, of imposing the contact line boundary condition to an unknown number and reconfigurable contact lines, when modeling the spreading dynamics on textured substrates, are now overcome. The validity of our approach is tested against experimental data of a droplet impacting on a horizontal solid surface. The study of the early spreading stage on hierarchically structured and chemically patterned solid substrates reveal an inertial regime where the contact radius grows according to a universal power law, perfectly agreeing with recently published experimental findings.

\end{abstract}

\maketitle
%
%
\section*{Introduction}
Recent applications of a droplet displacement in microfluidic devices (including smart optics\cite{karapetsas2011, You2013}, printing\cite{Wijshoff2010} and energy harvesting\cite{Krupenkin2011}) render the modeling of the contact line movement as a crucial research topic. Despite the huge amount of work published in the last thirty years, the appropriate boundary condition imposed at the contact line is yet a controversial issue since the governing physical processes are still unclear\cite{neto2005, Blake2006, lauga2007, bonn2009, Ren2010, lee2014}. In particular, both the two prevailing modeling approaches, namely the hydrodynamic model\cite{Huh1971} and the molecular kinetic theory\cite{Blake1969} can exhibit discrepancies from the experimental data\cite{Fetzer2009}, suggesting that a complete description of the physics is not yet provided. Worth mentioning the Huh and Scriven famous quote\cite{Huh1971}:``\textit{not even Herakles could sink a solid if the physical model were entirely valid, which it is not}'', referring to the contact line singularity arising at the hydrodynamic model when the Navier-Stokes equations along with the no-slip boundary condition at the solid surface are used.
\par
A slip boundary condition, where the tangential component of the velocity appears to have a finite value at the contact line, has been proposed in order to remove the hydrodynamic model singularity\cite{dussan1976, ren2007}. Common trait of these kind of approximations is the introduction of theoretical correlations between the contact line velocity and the dynamic contact angle (the angle formed between a droplet and the surface), imposed at the contact line (e.g. Cox formula\cite{Cox1986}, Hoffman-Voinov-Tanner law\cite{hoffman1975, voinov1976, tanner1979}, etc.). There is also the case where the value of the contact angle is fixed and equal to its static value (or Young's contact angle) as defined by the Young equation\cite{de1985}. Although these modified hydrodynamic models do succeed in removing the singularity arising in the motion of the contact line, they do not have general applicability since many physical quantities are explicitly imposed, though sometimes, derived from molecular dynamics simulations. In addition, in the case of modeling droplet dynamics on a geometrically patterned solid surface, the implementation of the contact angle boundary condition is not a trivial task since the change in the droplet topology demands a reconfiguration of the contact lines. As a result, the contact angle boundary condition has to be applied in \textit{a priori} unknown cardinality of contact lines hindering the simulation of the physical system. An attempt to overcome this restriction was performed by Savva and Kalliadasis \cite{savva2009} who studied the spreading of a droplet on a patterned substrate by:
\\a) utilizing the long-wave approximation of the Stokes equations and
\\b) treating the contact line as a set of two points, at the droplet edges.
\\The applicability of this methodology is however limited only for slow flows and small contact angles. In addition, the experimentally observed\cite{peters2009} entrapment of the ambient phase under the droplet is totally neglected.
\par
Another approach for modeling the contact line dynamics include fine-scale methods (e.g., molecular dynamics\cite{koplik1995, wang2013} or mesoscopic lattice Boltzmann models\cite{briant2004, kavousanakis2012, colosqui2013}). In contrast with the continuum-level models, fine-scale methods have led to better understanding of the wetting dynamics (e.g., Thompson and Robbins\cite{thompson1989} demonstrated that the no-slip boundary condition breaks down within a slip region around the contact line), however they suffer from severe computational limitations when realistic-sized systems (e.g. droplets with millimeter-sized radii) are examined. Alternatively, in order to minimize the computational demands, the continuum-level hydrodynamic model can be appropriately modified so as to account for the physical processes near the contact line. Such a study was performed by Ren and E\cite{ren2007}, where the boundary condition at the contact line is derived from systematic molecular dynamics simulations. Nevertheless, the above methodology is impracticable when studying wetting dynamics on complex (geometrically structured) surfaces due to the arduous implementation of the boundary condition.
\par
We recently presented static equilibrium computations of droplets with multiple contact lines, wetting geometrically patterned solid surfaces\cite{Chamakos2013, chamakos2014, Kavousanakis2015}. According to our approach, the liquid/ambient (LA) and the liquid/solid interfaces are treated in a unified context (one equation for both interfaces) 
by: \\a) employing the Young-Laplace equation\cite{adamson1990} augmented with a disjoining (or Derjaguin) pressure term\cite{deryaguin1987, Starov2010}, which accounts for the micro-scale liquid/solid interactions, and \\b) parameterizing the liquid surface in terms of its arc-length of the effectively one-dimensional droplet profile. \\The main advantage of this method
is that it avoids the implementation of any boundary condition at the contact line (the static, or Young, contact angle emerges implicitly as the result of the combined action of the disjoining and the capillary pressure) thus enabling the study of wetting phenomena on roughened solid substrates.
\par
In this work, we utilize this proposed liquid/solid interaction formulation, expressed by the disjoining pressure term to perform dynamic simulations of axisymmetric droplets spreading on solid substrates covered with arbitrary asperities, in air ambient. This modeling approach does not suffer from the nonphysical stress singularity at the contact line. In addition, contrary to the conventional hydrodynamic model, the relative velocity between the solid and the liquid is not directly imposed with a boundary condition (e.g. no-slip boundary condition), but it emerges implicitly as a result of the energy dissipation taking place on the intrinsic geometrical defects of the solid surface. Such an argument was introduced by Richardson\cite{richardson1973} and Jansons\cite{jansons1988} who concluded that even a small amount of roughness can approximate macroscopically a no-slip boundary condition, even when the solid material is microscopically perfectly slipping.  This assertion was later mathematically proved for a three dimensional flow by Casado-Diaz et al.\cite{casado2003}. The same applies for the dynamic contact angle which is also derived ``naturally'' in our formuation and not set by a specific boundary condition.
\par
The present article is organized as follows: we first present the mathematical framework of the dynamic problem to be solved. In the following section, the computational results (i.e. contact radius and dynamic contact angle of the droplet against time) are validated with experimental measurements and the influence of surface roughness geometric characteristics on the spreading dynamics is discussed. Next, computations of droplets spreading on chemically patterned and hierarchical roughened solid surfaces (where an intrinsic surface roughness is superimposed on larger structures) are performed in order to examine a recently proposed argument\cite{stapelbroek2014} of a universal droplet evolution regime at the early spreading stages, regardless of the underlying substrate. Concluding annotations are made in the last section.
%
%
\section*{Mathematical formulation}
\begin{figure}
\includegraphics[scale=0.7]{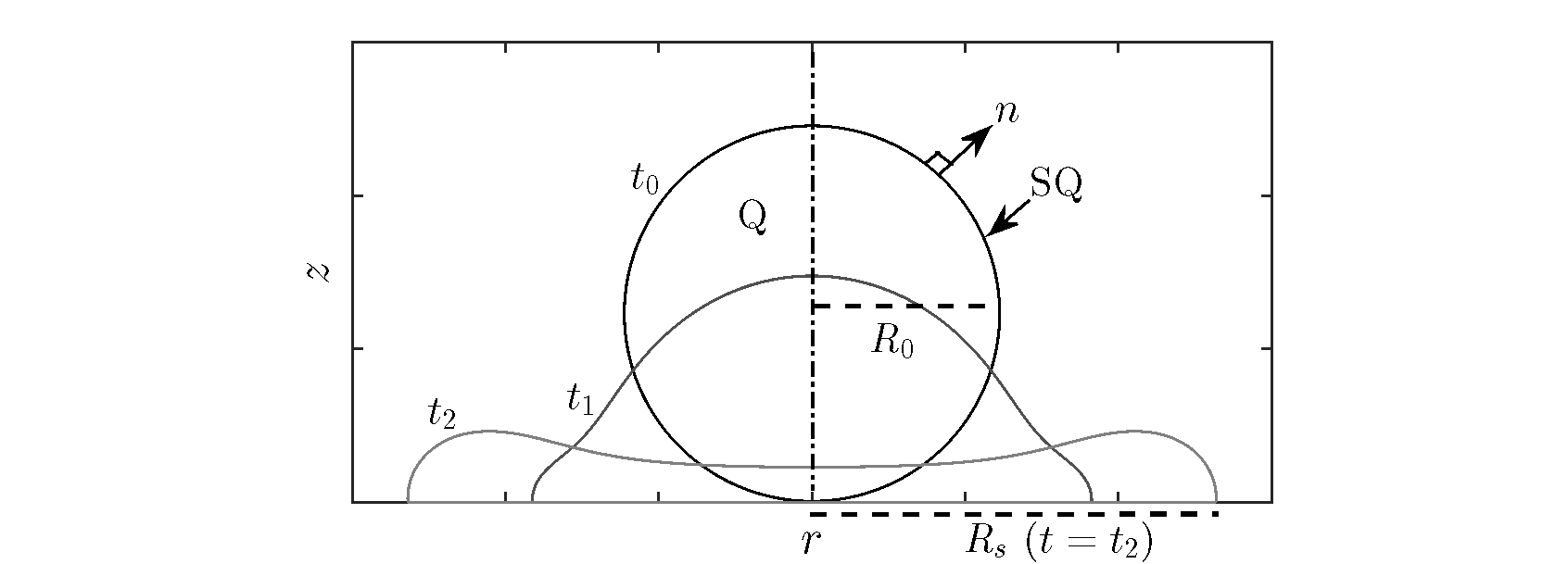}
\caption{Snapshots of an axisymmetric droplet at different time instances ($t_0 < t_1 < t_2$) after impacting on a flat solid substrate.}
\label{fig1}
\end{figure}

Since the viscosity of the surrounding phase (air ambient) in our system is negligible compared to that of the droplet, the flow problem is solved only for the droplet interior (Q in Fig.~(\ref{fig1})). In particular, we employ the incompressible Navier-Stokes equations\cite{landau2013} for Newtonian fluids in cylindrical coordinates $(r,z)$, to model the droplet spreading dynamics:
\begin{subequations}
\label{navierstokes}
\begin{eqnarray}
\rho(\partial_t{\boldsymbol{u}}+\boldsymbol{u} \cdot \nabla{\boldsymbol{u}})= - \nabla p + \mu \nabla^2 \boldsymbol{u} + \boldsymbol{F}, \label{navierstokes1}
\\
\nabla \cdot {\boldsymbol{u}}=0, {\text{in Q,}}  \label{navierstokes2}
\end{eqnarray}
\end{subequations}
where $\boldsymbol{u}$ is the fluid velocity field; $p$, $\mu$ and $\rho$ represent the pressure, dynamic viscosity and density of the fluid, respectively; $\boldsymbol{F}$ is a body force expressing the effect of gravity ($\boldsymbol{F}=\rho \boldsymbol{g}$, where the vector $\boldsymbol{g}$ is the gravitational acceleration). Equation~(\ref{navierstokes1}) states the momentum balance, and Eq.~(\ref{navierstokes2}) expresses the incompressibility of the droplet. The solution of the above set of equations is determined subject to a boundary condition, which is applied at the liquid/ambient interface (SQ in Fig.~(\ref{fig1})), and states the local force balance between surface tension, viscous stresses and liquid/solid interactions\cite{Scardovelli1999}:
\begin{equation}
\boldsymbol{n} \cdot \boldsymbol{\tau} = - p_{ext} \boldsymbol{n} -2 \gamma \kappa \boldsymbol{n}, {\text{at SQ.}}\label{navierbcondition}
\end{equation}
In the above, $\boldsymbol{n}$ is the outward unit normal of the liquid/ambient interface; $\boldsymbol{\tau}$ is the total stress tensor, $\boldsymbol{\tau} = -p \boldsymbol{I} + \mu (\nabla \boldsymbol{u} + (\nabla \boldsymbol{u} )^\mathrm{T}) $; $\gamma$ represents the surface tension of the liquid/ambient interface and $\kappa$ is the corresponding mean curvature. The latter is defined as:
\begin{equation}
\kappa= \frac{1}{2} \nabla_s \cdot \boldsymbol{n}, \label{mean_curvature}
\end{equation}
where $\nabla_s$ is the surface gradient operator ($\nabla_s=\nabla-\boldsymbol{n} \left( \boldsymbol{n} \cdot \nabla \right)$).
\par
Incorporating micro-scale interaction in the Navier-Stokes equations is not unusual in the literature (e.g. an extra body force, representing the van der Waals interactions, has been used in a plethora of publications studying liquid films rapture\cite{ruckenstein1974, williams1982}). Lubrication theory has also been used to study moving contact lines, taking into account the repulsive and attractive interactions\cite{eggers2005contact}. Despite the fact that in the latter case no boundary condition is required for the contact line (the liquid tends to a precursor film), it is not applicable for complex solid topographies. In our case we employ micro-scale liquid/solid interactions, lumped in a disjoining pressure term, in order to stabilize an intermediate thin layer between the liquid and solid phases. The above interactions are imposed via a pressure term, $p_{ext}$, which appears in the interface force balance (Eq.~(\ref{navierbcondition})). Specifically, we adopt the following formulation:
\begin{equation}
p_{ext}=p_0+p^{LS},\label{externalpressure}
\end{equation}
where $p_0$ is a reference pressure (constant along the interface) representing the pressure of the ambient phase. The disjoining pressure term, $p^{LS}$, expresses the excess pressure due to the liquid/solid interactions  and is given by the following expression\cite{Chamakos2013, chamakos2014}:
\begin{equation}
p^{LS}={{\gamma}\over {R_0}}w^{LS}\left[ \left( {{\sigma} \over {\delta/R_0+\epsilon}} \right)^{C_1}-\left( {{\sigma} \over {\delta/R_0+\epsilon}} \right)^{C_2} \right], \label{disjoiningpressure}
\end{equation}
which essentially resembles a Lennard-Jones\cite{atkins2014} type potential. In particular, the depth of the potential well is proportional to the wetting parameter, $w^{LS}$, which is directly related with the solid wettability (an increase of $w^{LS}$ results in a deeper well of the potential, indicating stronger liquid/solid interaction). The exponents ${C_1}$ and ${C_2}$ control the range of the micro-scale liquid/solid interactions (large ${C_1}$ and ${C_2}$ reduce the range within which micro-scale interactions are active) and $R_0$ is a characteristic length (here the initial radius of the spherical droplet). The distance, $\delta$, between the liquid and the solid surfaces determines whether the disjoining pressure is attractive (modeling van der Waals interactions, for relatively large $\delta$) or repulsive (modeling the overlapping of the electrical double layers, for small $\delta$). In the case of a perfectly flat solid surface, the distance $\delta$ is defined as the vertical distance of the liquid surface from the solid boundary. For non-flat, rough, solid surfaces, the definition of distance, $\delta$, requires special consideration. Here, we take $\delta$ as the Euclidean distance from the solid. This quantity is obtained from the solution of the Eikonal equation\cite{dacorogna1999} which expresses the signed distance from a boundary (even arbitrarily shaped), as proposed in our previous work\cite{Chamakos2013, chamakos2014}. The solution of the Eikonal equation is discussed in Appendix A. 
In our formulation the droplet never actually touches the solid wall since we assume that the liquid and the solid phases are separated by an intermediate layer (with thickness $\delta_{min}$). This intermediate layer is stabilized by the presence of the disjoining pressure. In particular at $\delta = \delta_{min}$ the repulsive and attractive micro-scale forces balance each other; further reduction of the intermediate layer thickness, below $\delta_{min}$, would generate strong repulsion. The minimum allowed liquid/solid distance $\delta_{min}$ is controlled by the constants ${\sigma}$ and $\epsilon$.
Specifically, for $\delta = \delta_{min} \Leftrightarrow p^{LS} =0 \Rightarrow \delta_{min} = R_0 (\sigma-\epsilon)$.
The selection of the disjoining pressure constants value is in accordance with our previous work\cite{chamakos2014}, namely: ${C_1}$ = 12, ${C_2}$ = 10, ${\sigma}=9 \times 10^{-3}$ and $\epsilon=8 \times 10^{-3}$.
A sensitivity analysis regarding the effect of disjoining pressure parameters on the spreading dynamics, as well the visualization of the corresponding $p^{LS}$ profiles, are presented in Appendix B.
\par
Apart from the Navier-Stokes equations (Eq.~(\ref{navierstokes})), the mesh smoothing equations need to be solved in order to efficiently account for the movement of the droplet surface.
The evolution of the liquid/ambient interface, which is advected by the velocity field, is defined by the following kinematic condition:
\begin{equation}
(\boldsymbol{u}_{mesh} - \boldsymbol{u}) \cdot \boldsymbol{n} = 0, {\text{at SQ,}}\label{meshvelocity}
\end{equation}
where $\boldsymbol{u}_{mesh}$ is the velocity of the mesh at the interface. The boundary displacement is therefore propagated throughout the domain resulting in a distortion of the computational mesh.  Indicatively, the unstructured mesh - delimited by the droplet surface - is appropriately deformed according to the Winslow elliptic smoothing equations\cite{Knupp1999}. In addition, the grid is regenerated when the interface undergoes large deformations. Details of the Winslow equations solution procedure are provided in Appendix C.
\par
Since the Young contact angle, $\theta_Y$, is not imposed explicitly in the simulation (both liquid/ambient and liquid/solid interfaces are treated in a consolidated framework), a correlation between the liquid/solid affinity, expressed via the wetting parameter, $w^{LS}$, and Young's contact angle is required. This is derived using a Frumkin-Derjaguin type formula\cite{Chamakos2013, chamakos2014, rauscher2008}:
\begin{equation}
cos\theta_Y = {{\omega_{min}} \over {\gamma}} - 1, \label{frumkin-derjaguin}
\end{equation}
where $\omega_{min}$ is the absolute minimum value of an effective interface potential, $\omega$, which essentially expresses the cost of free energy per unit area to maintain an intermediate layer, with thickness $\delta$, between the liquid and solid phases ($\omega \rightarrow 0 $ when $\delta \rightarrow \infty$). The effective interface potential is related with the disjoining pressure according to\cite{rauscher2008}:
\begin{equation}
p^{LS}=-{{{\text{d}}\omega}\over{{\text{d}}\delta}}, \label{disjoingpressure-omega}
\end{equation}
and reaches its minimum value, $\omega_{min}$, at the distance $\delta_{min}$ where $p^{LS}$ = 0. Equation~(\ref{frumkin-derjaguin}) is used to construct an one-to-one correlation between the Young contact angle and the wetting parameter ($w^{LS}=w^{LS}(\theta_Y)$) which reads:
\begin{equation}
w^{LS} =\frac{(C_1-1) (C_2-1) (1+cos\theta_Y)}{ \sigma (C_1-C_2) }. \label{thy-wls_correlation}
\end{equation}
The results of Eq.~(\ref{thy-wls_correlation}) can also be validated by performing a circular fitting on the computed free surface of the droplet profile at equilibrium\cite{Chamakos2013, chamakos2014}.
\par
Overall, the Navier-Stokes equations (Eq.~(\ref{navierstokes})) and the Winslow equations (see Appendix C) are discretized using the finite element method (FEM)\cite{zienkiewicz1971} accounting for the boundary conditions~(\ref{navierbcondition}) and~(\ref{meshvelocity}). The resulting set of discrete equations is integrated in time using the implicit Euler method in order to simulate the droplet impact and spreading on a solid substrate. By assuming axial symmetry, the model becomes two-dimensional (see Fig.~(\ref{fig1})) and is implemented in COMSOL Multiphysics\textsuperscript{\textregistered} commercial software.
%
%
\section*{Results and discussion}
%
%
\subsection{Spreading dynamics - Validation with experimental data}
The proposed ``boundary-condition-free'' approach is implemented to model the dynamics of a droplet impacting and spreading on a horizontal surface. In order to validate our predictions, the results are compared with experimental data\cite{sikalo2005} of a glycerin/water mixture droplet (85 \% of glycerin) spreading on a wax solid surface. In the studied experiment, the droplet impacts and then spreads on a solid surface with an initial vertical speed of $u_0$ = 1.04 m/s. This corresponds to a Weber number, $We$ = 51.2, and a Reynolds number, $Re$ = 26.8 ($We={{\rho {u_0}^2 2 R_0} \over {\gamma}}$ and $Re={{\rho u_0 2 R_0} \over {\mu}}$, where $\rho$ = 1220 kg/m\textsuperscript{3}, $R_0$ = 1.225 mm, $\gamma$ = 0.063 N/m and $\mu$ = 116 mPa s). The gravitational force, $\boldsymbol{F}$, in the Navier-Stokes equations (Eq.~(\ref{navierstokes})) is neglected since the initial droplet radius, $R_0$, is smaller than the capillary length, $\lambda_{\kappa}$ ($\lambda_{\kappa} = \sqrt{{\gamma\over{\rho g}}}$ = 2.294 mm for the used glycerin/water mixture droplet). Finally, the wettability of the wax substrate corresponds to a Young contact angle, $\theta_Y$ = 93.5\textsuperscript{o}, which is equivalent to a wetting parameter $w^{LS} = 5.16 \times 10^{3}$, given from  Eq.~(\ref{thy-wls_correlation}).
\begin{figure}
\includegraphics[scale=0.7]{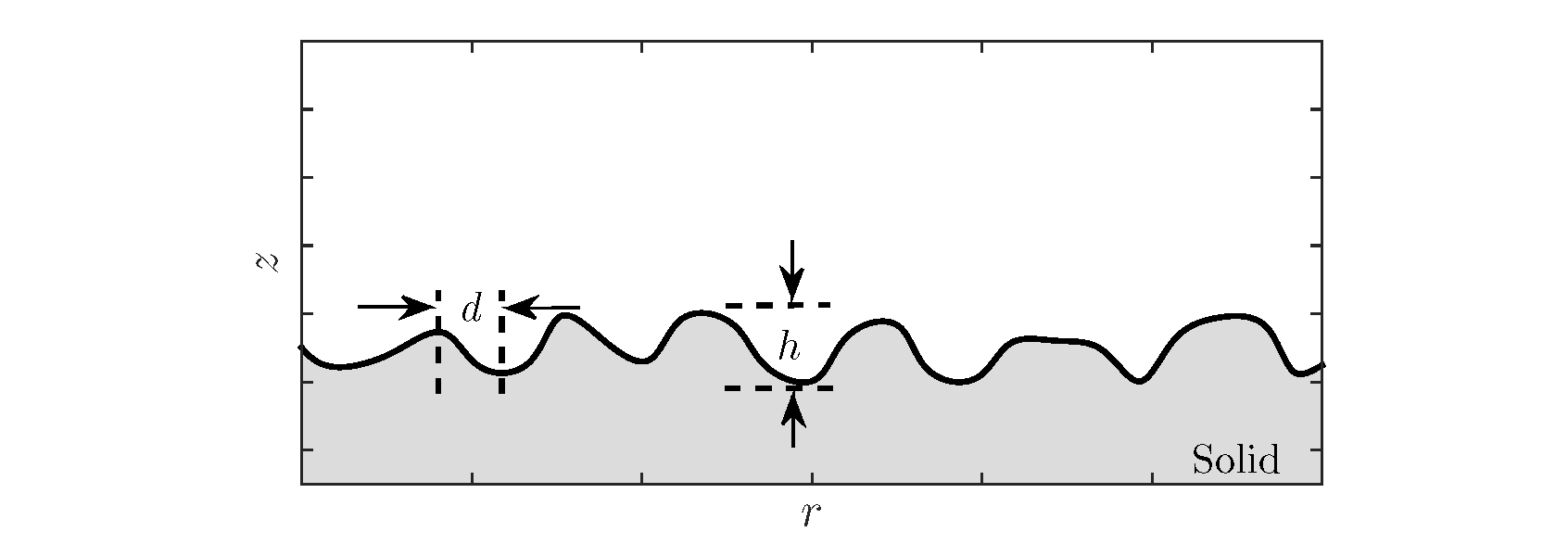}
\caption{Geometric parameters of the solid surface intrinsic roughness.}
\label{fig2}
\end{figure}
\par
There are previous studies\cite{sikalo2005_2, lunkad2007} where the conventional hydrodynamic model has been utilized to simulate the same experiments. Despite the fact that their results do succeed in capturing some of the experimental trends, the above approaches are case sensitive since {\it{ad hoc}} correlations are used to model the dynamic contact angle and the shear stresses at the contact line. In particular, in the first study\cite{sikalo2005_2} Sikalo et al. use a semi-empirical correlation given by Kistler\cite{Kistler_1993}, whereas in the latter\cite{lunkad2007} a time variation of the contact angle, based on experimental measurements, is employed. Here, by using the suggested formulation, the effective shear stresses arise macroscopically due to the micro-scale roughness of the solid surface (note that disjoining pressure induces only normal forces to the liquid/ambient interface (see Eq.~(\ref{navierbcondition}))). Aiming to mimic the intrinsic roughness of the wax surface, we employ an arbitrary roughness topography, which is characterized by two length parameters: a maximum amplitude of the protrusions, $h$, and an average distance between two neighboring extrema of the surface, $d$ (see Fig.~(\ref{fig2})). The solid surface topography is produced with the following procedure: Initially, an array of random scalars, representing the local extrema of the substrate profile, is drawn from the standard uniform distribution on the interval $(0,h)$ and the distance between them is set equal to $d$. Next, the final form of the solid topography is obtained by performing a piecewise cubic interpolation\cite{fritsch1980} on the random extrema of the substrate, ensuring that the substrate profile and its derivatives are continuous. In order to quantify the non-uniform, arbitrary pattern of the solid protrusions we also introduce the surface roughness factor, $r_f$, which is defined as the ratio of the actual over the apparent surface area ($r_f\ge 1$).
The effect of the roughness factor on the spreading dynamics is examined in the following section.
%
%
\subsubsection{The effect of the micro-scale roughness factor}
In the cases under study, the intrinsic roughness factor of the modeled surface varies from $r_f$ = 1 (for an ideally smooth surface) to $r_f$ = 1.2 (for the most roughened solid surface). In the following sections, when referring, for brevity, to a ``smooth" surface we mean an ideally smooth and flat substrate. The solid surface rugosity is parametrized by the average distance of the inhomogeneities, $d$, (0.012 mm $\le d \le$ 0.06 mm), and by their maximum height, $h$, (0.012 mm $\le h \le$ 0.055 mm). Note that the ratio of the smallest roughness height ($h$ = 0.012 mm) over the initial droplet diameter, $2 R_0$, is $\frac{1}{204}$. 
\par
\begin{figure}
\includegraphics[scale=0.7]{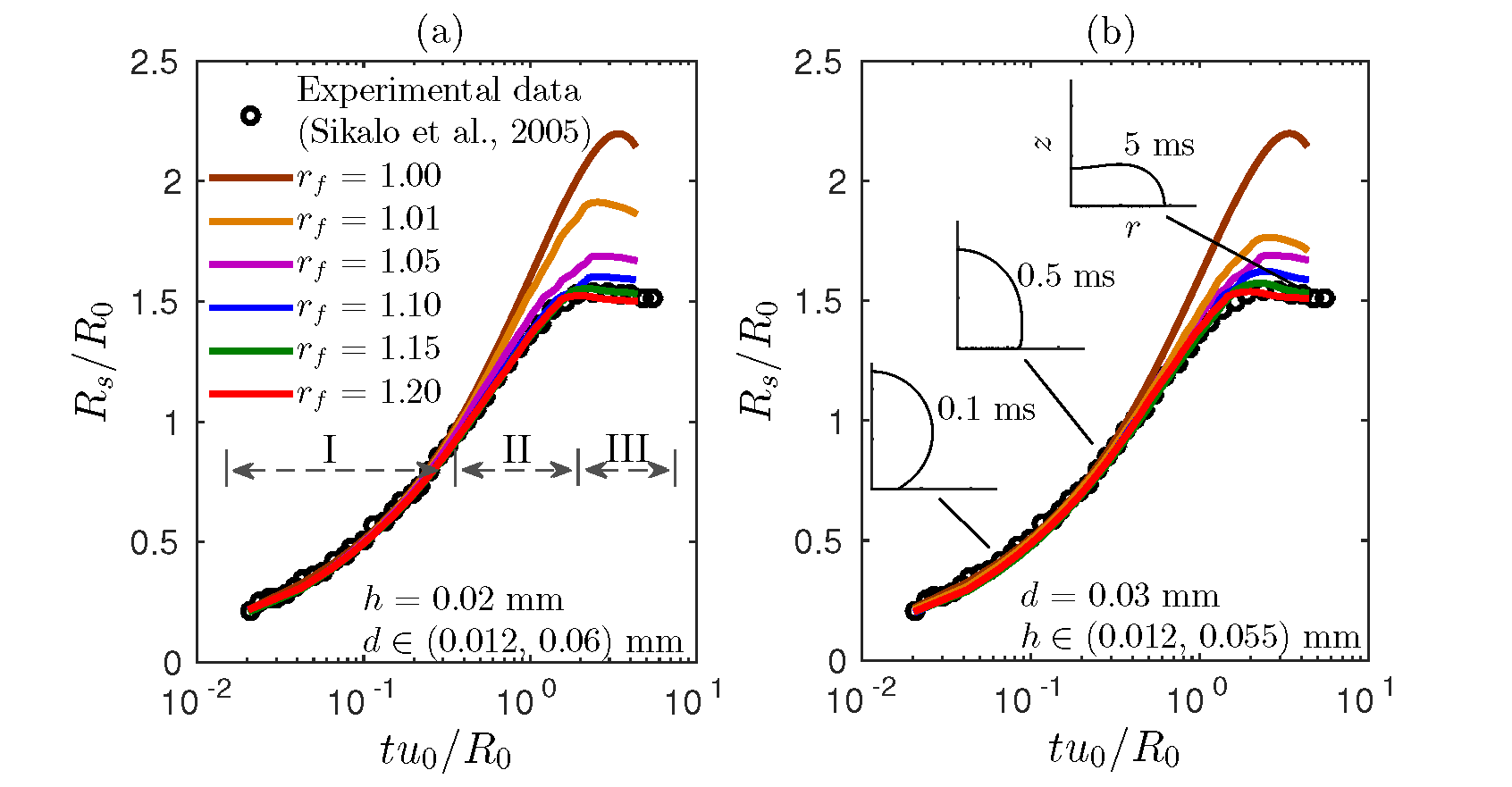}
\caption{Temporal evolution of the normalized contact radius of a glycerin/water mixture droplet impacting on wax surface: simulations against experimental data from Sikalo et al.\cite{sikalo2005} for different roughness factor cases. 
The roughness is increased by: (a) reducing the average distance of the inhomogeneities, $d$, or (b) enlarging the protrusions amplitude, $h$.}
\label{fig3}
\end{figure}

In Fig.~(\ref{fig3}), we present the temporal evolution of the normalized contact radius, $R_s/R_0$ (see also Fig.~(\ref{fig1})), for different solid surface roughness cases. In order to evaluate the contact radius we consider that the contact line is defined as the intersection of the droplet surface with a horizontal baseline just above the substrate ($z \approx 4 \times 10^{-3} R_0$). The initial frame, i.e. $t=0$, is the instance of the droplet impact. The time is presented in dimensionless form and the characteristic time is $t u_0/ R_0$.
The surface roughness factor, here, is increased either by decreasing the average distance between the surface inhomogeneities while keeping fixed their maximum amplitude (see Fig.~(\ref{fig3}a)), or by increasing their maximum amplitude by fixing their distance constant (see Fig.~(\ref{fig3}b)).
\par
Our results show that the spreading behavior is not similar for the different surface roughness cases. The spreading evolves in three stages (see Fig.~(\ref{fig3})): 
\\(I) An early spreading stage, where the radius sharply increases with time (notice the concave part of the curve). At this stage the spreading evolution does not depend on the roughness factor.
\\(II) An intermediate stage, where an almost linear dependence of the normalized contact radius on the logarithmic characteristic time can be observed. At the end of this stage the effect of the roughness factor is visible. As it can be seen, the more roughened the substrate, the slower the contact radius evolution.
\\(III) A recoiling stage, where the droplet begins to recede and the contact radius slightly reduces after reaching a maximum value. This effect is particularly visible in the case of the perfectly smooth surface. As it can be observed, at the recoiling stage, the positions of the contact line for the smooth and the rough substrate cases significantly differ.
\\Interestingly enough, the computational results converge to the experimental data, regardless of the geometric details of the substrate, above a certain roughness factor threshold (here for $r_f>$ 1.15). The negligible effect of the exact roughness topography (for $r_f>$ 1.15) on the impact dynamics, is an indication that the assumption of the axially-symmetric substrate does not play a key role in the obtained results.
\par
\begin{figure}
\includegraphics[scale=0.7]{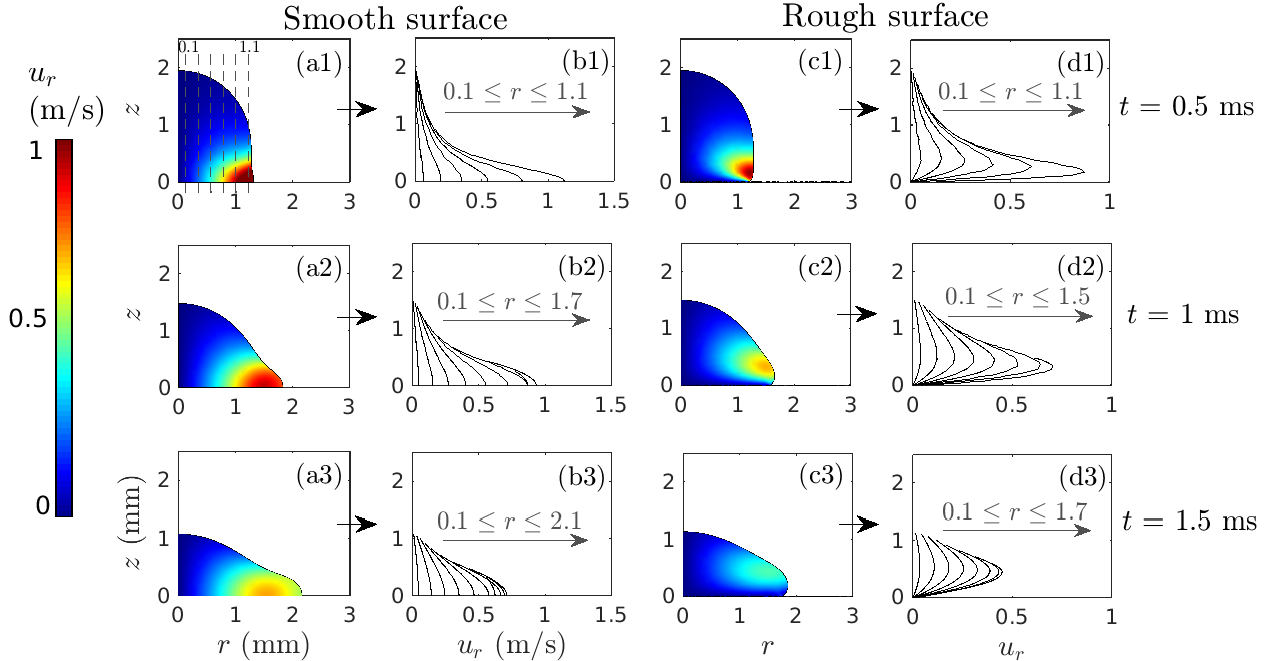}
\caption{Visualization of the $r$-component of the droplet velocity, $u_r$, for different roughness factor cases: (a) $r_f$ = 1 and (c) $r_f$ = 1.2, at various time instances ($t \in$ [0.5 ms, 1.5 ms]).
The $u_r$ distribution is also plotted along vertical, equidistant cut lines (with a $\delta r$ step of 0.2 mm) for (b) smooth and (d) rough substrate cases.}
\label{fig4}
\end{figure}
Aiming to unravel the effect of the solid substrate roughness on the flow dynamics, we visualize in Fig.~(\ref{fig4}) the $r$-component of the velocity, $u_r$, for a droplet spreading on a smooth and a rough solid surface. By plotting the $u_r$ distribution along equidistant cut lines of constant $r$ (see Fig.~(\ref{fig4}d1-d3)), it can be observed that the velocity considerably decreases and reaches almost zero, especially very close to the substrate (for $z \rightarrow 0$), in the case of the rough surface. This indicates that small variations in the velocity direction, as the fluid slips on the solid asperities, dissipates a considerable amount of energy. Clearly, the total dissipation in the roughness vicinity can approximate a no-slip boundary condition, without imposing directly any shear interaction between the liquid and the solid surfaces. On the contrary, in the absence of an intrinsic surface roughness, the velocity has a finite value on the wall since the energy dissipation is negligible (see Fig.~(\ref{fig4}b1-b3)). Note that such a free-slip case is non-realistic since a microscopic slip length has been observed (in molecular dynamics simulations) even for molecularly smooth surfaces due to the discrete nature of the lattice structure of the solids\cite{koplik1989molecular}. Here, however, we use this case in order to highlight that the no-slip boundary condition can occur as a solely geometric effect, even at a molecular level. The above is in line with the work of Richardson\cite{richardson1973} and Jansons\cite{jansons1988}.
\begin{figure}
\includegraphics[scale=0.7]{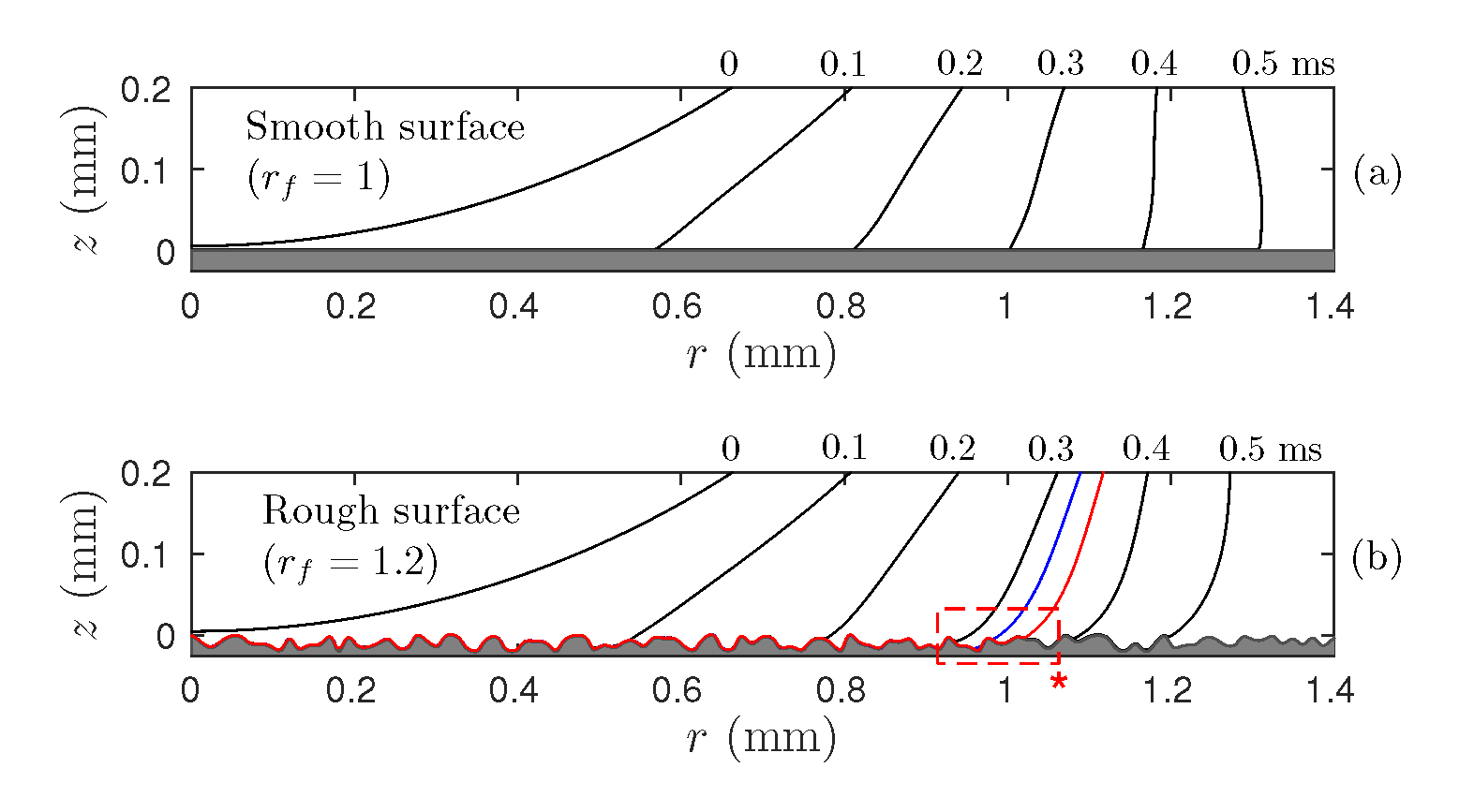}\\
\includegraphics[scale=0.8]{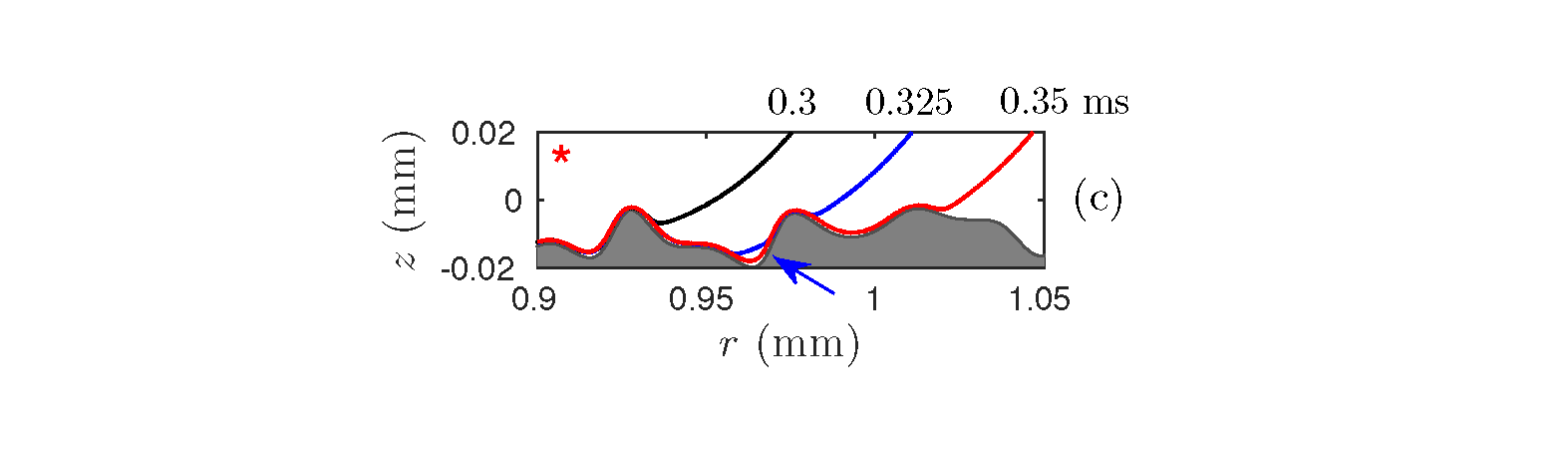}
\caption{Droplet profiles in the vicinity of: (a) a perfectly smooth ($r_f$ = 1) and (b) a rough solid surface ($r_f$ = 1.2) during the initial spreading stages ($t\le$ 0.5 ms). A magnification of the droplet shape, for the area outlined in (b), is depicted in (c) for $t \in$ [0.3 ms, 0.35 ms]; changes in the droplet topology (marked by the arrow) can be effectively handled by using our proposed formulation.}
\label{fig5}
\end{figure}
\par
\begin{figure}
\includegraphics[scale=0.7]{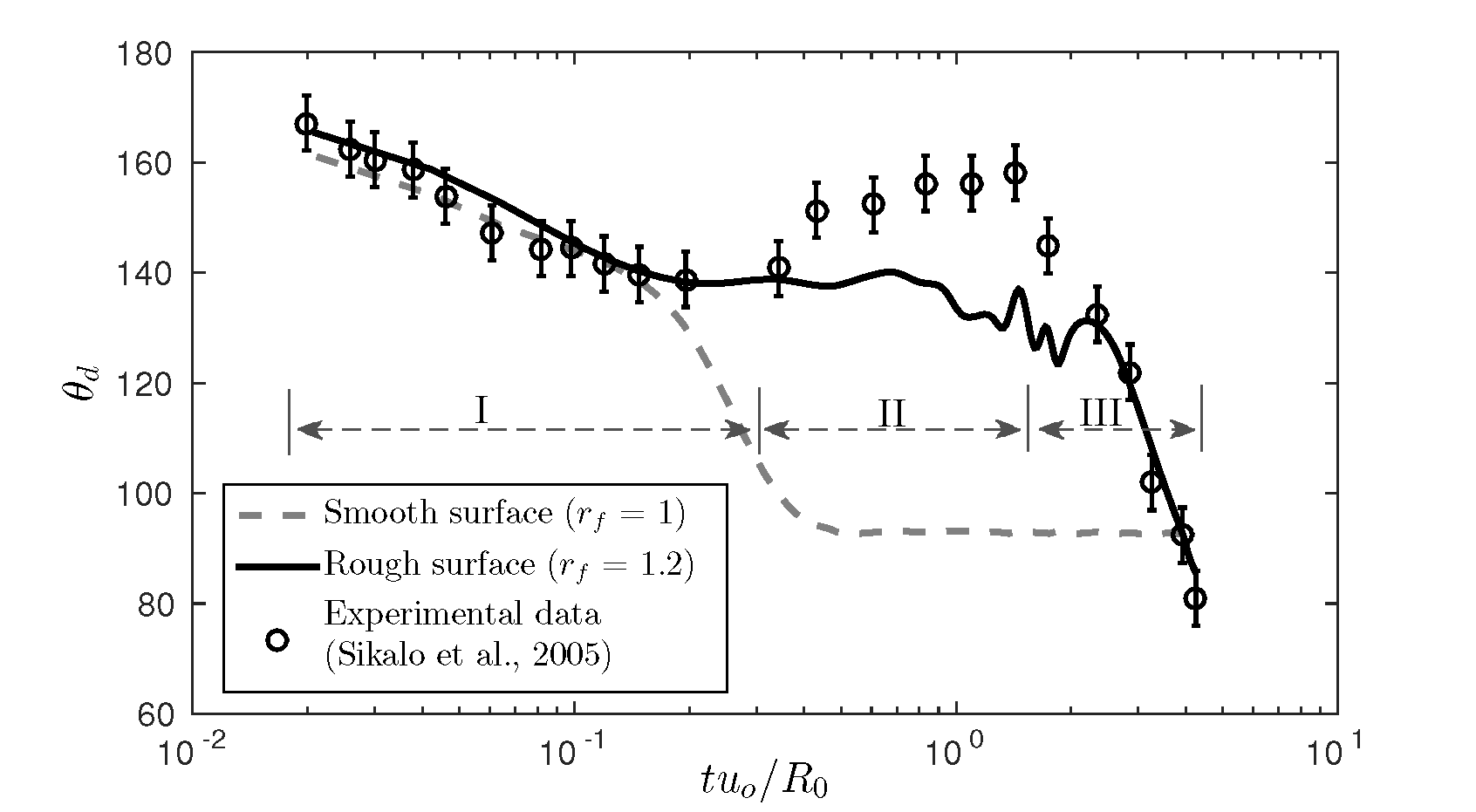}
\caption{Dynamic contact angle, $\theta_d$, of a glycerin/water mixture droplet impacting on wax surface: Computations, on a smooth ($r_f$ = 1) and a rough ($r_f$ = 1.2) solid substrate, are tested against experimental data from Sikalo et al.\cite{sikalo2005}}
\label{fig6}
\end{figure}

A magnification of the droplet profile in the vicinity of the solid surface, shown in Fig.~(\ref{fig5}), provides a detailed view of the droplet deformation during the initial spreading stages ($t<$ 0.5 ms). The discrepancy between the contact radius at the smooth (Fig.~(\ref{fig5}a)) and the rough solid surface (Fig.~(\ref{fig5}b)) is visible, especially on the last frame ($t$ = 0.5 ms). The inherit capability of our formulation to handle changes in the droplet topology is presented in Fig.~(\ref{fig5}c). Formation ($t$ = 0.325 ms) as well as destruction ($t$ = 0.35 ms) of contact lines can be observed as the droplet slips on the corrugated substrate. Regarding the evolution of the contact angle, it is noticeable that the two cases (perfectly smooth and roughened substrate) exhibit large differences. We note here that despite the fact that in the conventional hydrodynamic models the apparent (or macroscopic) dynamic contact angle is set as a function between the equilibrium (Young's) contact angle and the capillary number, $Ca$ ($Ca = u_{cl} \mu / \gamma $, where $u_{cl}$ is the fluid velocity magnitude at the contact line), in our modeling approach, it emerges ``naturally" as a result of the local interplay of viscous, capillary and liquid/solid interaction forces. Indicatively, in Fig.~(\ref{fig6}) we present the dynamic contact angle value, ${\theta_d}$, of the droplet for a smooth ($r_f$ = 1) and a rough ($r_f$ = 1.2) solid surface cases, as a function of time. Unfortunately, as Sikalo et al.\cite{sikalo2005} claim, the dynamic contact angle measurements are at limited accuracy (the measured values depend on the experience of the experimentalist). Here, we apply their measurement technique in our computational predictions (namely to obtain the contact angle from the computed droplet profiles) in order to directly compare them with the experimental measurements. In particular, the slope of the droplet surface is evaluated, against the horizontal plane, at a fixed distance from the substrate ($z \approx 8 \times 10^{-3} R_0$) where the action of the disjoining pressure has been effectively vanished (region III in Fig.~(\ref{fig13}b), Appendix B).
\par
Our results in Fig.~(\ref{fig6}) show that the early spreading (I) and the recoiling (III) stages are adequately captured, for the case of rough solid substrate. During the intermediate stage (II), however, our results predict fluctuations around a plateau value of the contact angle which can be attributed to the contact line pinning-depinning on the substrate corrugations. This contrasts to Sikalo et al.\cite{sikalo2005} measurements, where a local maximum of the dynamic contact angle is observed. To our opinion the discrepancy is caused by the relative large height of the roughness protrusions, compared to the droplet size, which highly distorts the droplet shape at the contact line. Further decrease of the protrusions length scale, resulting in more realistic roughness structures (e.g. the documented average roughness amplitude for the wax surface is $0.3 \times 10^{-3}$ mm\cite{sikalo2005}), although it is feasible in our formulation, it requires extensive computational resources sacrificing the efficiency of the continuum-level modeling. Such a study is beyond the scope of this particular work. When the substrate is ideally smooth, the dynamic contact angle coincides with the Young contact angle ($\theta_d \approx \theta_Y$ = 93.5\textsuperscript{o}) shortly after the droplet impact, indicating again that an amount of roughness is essential in our modeling approach.
\par
Aiming to obtain more insight of the viscous effects resisting the droplet spreading, we next investigate the dissipation of energy during the droplet impact. In particular, we focus on the differences at the energy loss between the case of the ideally smooth and a roughened solid substrate.
%
%
\subsubsection{Quantifying the energy dissipation effect at the contact line}
The energy dissipation can be quantified by calculating the viscous dissipation function, $\phi$\cite{warsi2005}:
\begin{equation}
\phi = \boldsymbol{\tau} : \nabla  \boldsymbol{u}. \label{energy_dissipation}
\end{equation}
By evaluating the double colon product and utilizing the symmetry arguments, Eq.~(\ref{energy_dissipation}) in cylindrical coordinates reads:
\begin{equation}
\phi = {2 \mu} \left[ \left(\frac{\partial u_r}{\partial r}\right)^2 + \left(\frac{\partial u_z}{\partial z}\right)^2 +  \left(\frac{u_r}{r}\right)^2  + \frac{1}{2}\left(  \left(\frac{\partial u_z}{\partial r}\right)^2+ \left(\frac{\partial u_r}{\partial z}\right)^2 +2\frac{\partial u_z}{\partial r} \frac{\partial u_r}{\partial z}  \right)   \right], \label{energy_dissipation2}
\end{equation}
where $ u_z$ represents the $z$-component of the fluid velocity. 
\begin{figure}
\includegraphics[scale=0.7]{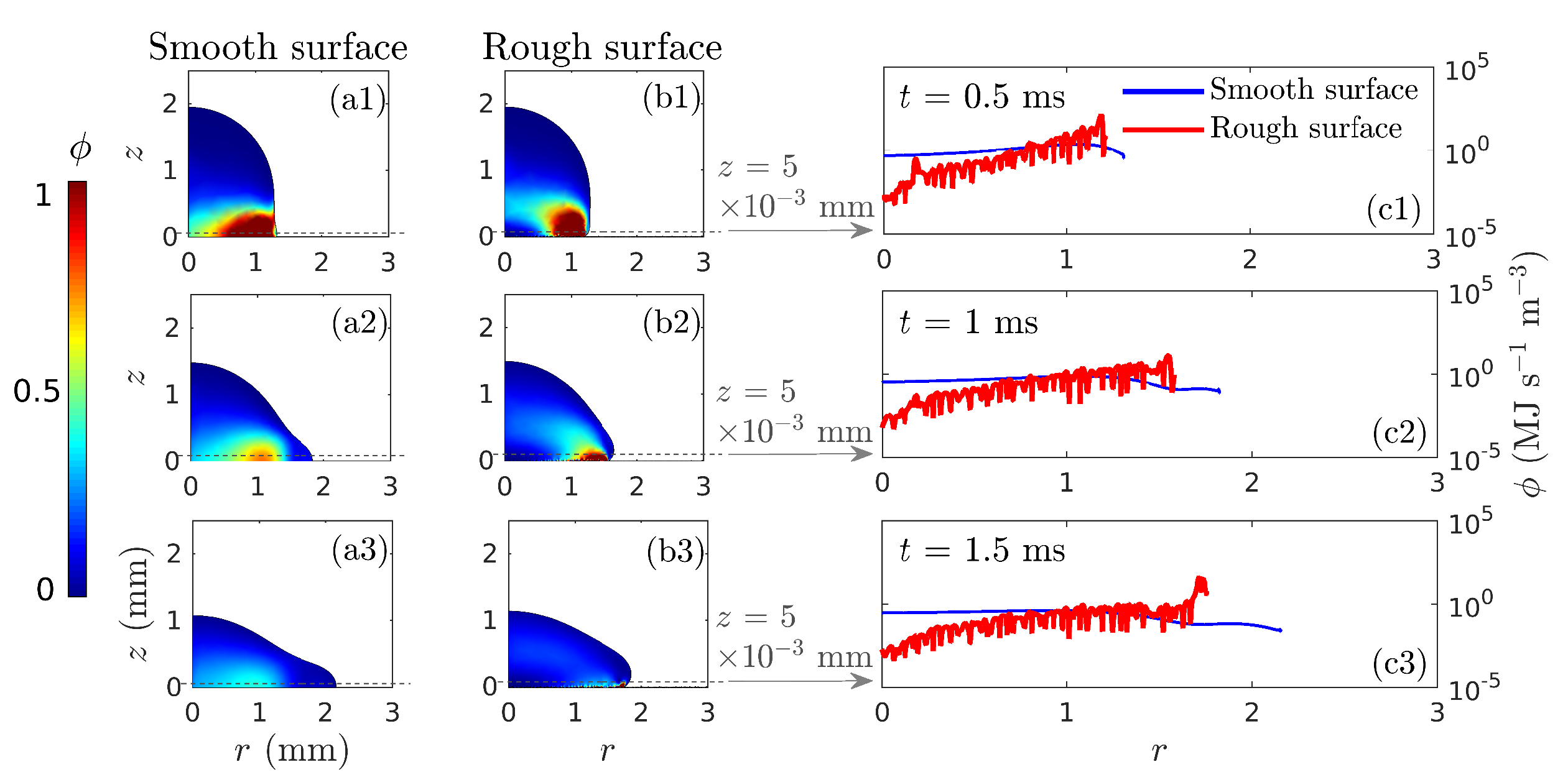}
\caption{Rate of energy dissipation per unit volume, $\phi$, during the impact process on: (a) a smooth solid surface ($r_f$ = 1) and (b) a rough solid surface ($r_f$ = 1.2), at different time instances ($t \in$ [0.5 ms, 1.5 ms]).
The distribution of $\phi$ along a line with constant height ($z = 5 \times 10^{-3}$ mm) above the substrate is depicted in the right panel of the figure ((c)).}
\label{fig7}
\end{figure}
In Fig.~(\ref{fig7}), we plot the spatial distribution of $\phi$, which practically expresses the rate of energy dissipation per unit volume of the fluid. For the better illustration of the dissipation rate, $\phi$ is plotted within a short range of values (from 0 to 1) in  Fig.~(\ref{fig7}a1-a3) and  Fig.~(\ref{fig7}b1-b3). It is observed that during the spreading process, the initial droplet energy in both the smooth and the rough solid surface cases is mainly dissipated at the proximity of solid boundary. By examining the variation of $\phi$ along a line with constant height $z = 5 \times 10^{-3}$ mm (see Fig.~(\ref{fig7}c1-c3)), we detect a substantial energy dissipation, in the case of the rough surface, close to the contact line (note that the $\phi$ values are presented on a logarithmic scale on the right panel of Fig.~(\ref{fig7})). This means that the energy loss, responsible for the spreading deceleration in the rough solid surface case, predominantly occurs at the contact line region as a fluid passes over the substrate irregularities. The fluctuations, observed in dissipation function distribution, correspond to the highly inhomogeneous geometrical features of the substrate. Away from the contact line, as observed in Fig.~(\ref{fig7}), the rate of dissipation is negligible since the liquid is effectively brought to rest. When the droplet is spreading on a perfectly smooth and slippery substrate, the energy dissipation is attributed only to the inertial forces resisting the fluid movement. In this case, as demonstrated in Fig.~(\ref{fig7}), the dissipation function, $\phi$, does not exhibit extreme values along the line parallel to the solid surface at height $z = 5 \times 10^{-3}$ mm, since the effect of the solid substrate is negligible. 
\begin{figure}
\includegraphics[scale=0.7]{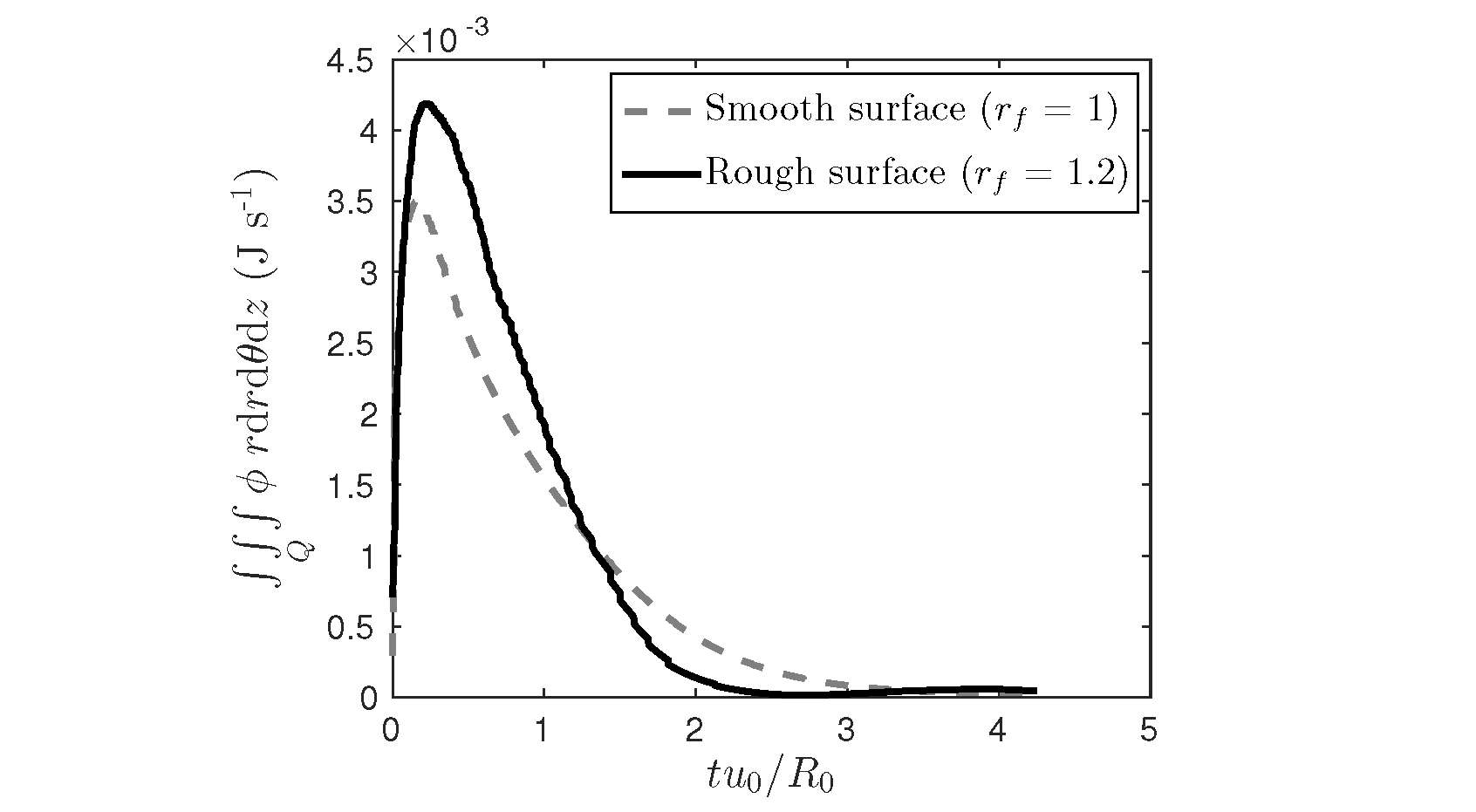}
\caption{Volume integral of the dissipation function in cylindrical coordinates, $\iiint\limits_Q \! \phi \, \mathrm{d}V = \iiint\limits_Q \! \phi \, r \mathrm{d} r \mathrm{d} \theta \mathrm{d} z$, during the spreading process for a perfectly smooth ($r_f$ =1) and a rough ($r_f$ = 1.2) solid surface case. In the expression of the volume element, $r \mathrm{d} r \mathrm{d} \theta \mathrm{d} z$, $\theta$ represents the cylindrical azimuthal coordinate ($\theta \in [0, 2\pi$)).}
\label{fig8}
\end{figure}
\par
The integration of $\phi$ over the entire volume of the droplet, $V$, presented in Fig.~(\ref{fig8}), also highlights the differences in the evolution of energy dissipation, for the rough and the smooth substrates, respectively. Specifically, for the rough solid surface case the overall rate of dissipation is high during the early spreading stages ($t u_0 / R_0 < 1$) followed by a sharp decrease. In the case of smooth surface the energy dissipation is milder. Although the rate of dissipation over time differs significantly between the two cases, a question arises about the relative ratio of the total amount of energy dissipated. Up to the time where the dissipation rate asymptotically gets very small (at $t u_0 / R_0 \approx 4$ or $t = 4.7$ ms) we calculate that the total energy consumed is: $E_d = \int_0^{4.7 \mathrm{ms}}  \iiint\limits_Q \! \phi \, \mathrm{d}V \mathrm{d}t = 4.72 \times 10^{-6}$ J for the rough substrate, and $E_d = 4.35 \times 10^{-6}$ J for the smooth substrate. The first corresponds to the 75.3 \% of the initial droplet energy before impact, $E_t$, and the latter to the 69.3 \% respectively, where $E_t$ is calculated as:
\begin{equation}
E_t = E_k + E_s = \frac{2}{3}\pi R_0^3 u_0^2 + 4 \pi R_0^2 \gamma, \label{droplet_energy}
\end{equation}
with $E_k$ the initial kinetic energy and $E_s$ the initial surface energy, respectively. From the above, we conclude that although the overall energy loss is virtually the same, the major differences observed in the spreading behavior on a rough and on a smooth solid surface, are mainly attributed to the locally dissimilar distribution of $\phi$ close to the contact line. 
\par
We believe that such an analysis can provide important information for the understanding of energy dissipating phenomena taking place at the contact line and would contribute in extending our understanding regarding the underlying complex physical mechanism. Finally, we note that the corresponding experimental measurement, of the dissipated energy during the spreading process, is until now unfeasible.
%
%
\subsection{Early spreading universality on complex surfaces}
The proposed modeling approach can be used to examine the effect of any kind of complex geometrical structure of the substrate (even hierarchical patterned solid surfaces) on the spreading dynamics - an arduous task for the conventional hydrodynamic models. Early time spreading dynamics phenomena, on partially or fully wetted substrates, are far from being fully understood\cite{yarin2006}, since the droplet behavior is affected by inertia, viscous and contact line friction effects; for later times, the energy dissipation mechanism is simplified and the dynamics can be adequately described by the well-verified Tanner's law\cite{tanner1979}, which is also recovered by our model (see Appendix D).
Stapelbroek et al. in a recent work\cite{stapelbroek2014} argued that the initial spreading dynamics of low viscosity fluids obey to a universal power law, independently of the geometric or chemical complexity of the solid substrate. Aiming to verify this argument, of the negligible role of the substrate during the early spreading stage, we next perform droplet impact computations on different types of solid surfaces, featuring topographical, as well as chemical heterogeneities.
\par
\begin{figure}
\includegraphics[scale=0.7]{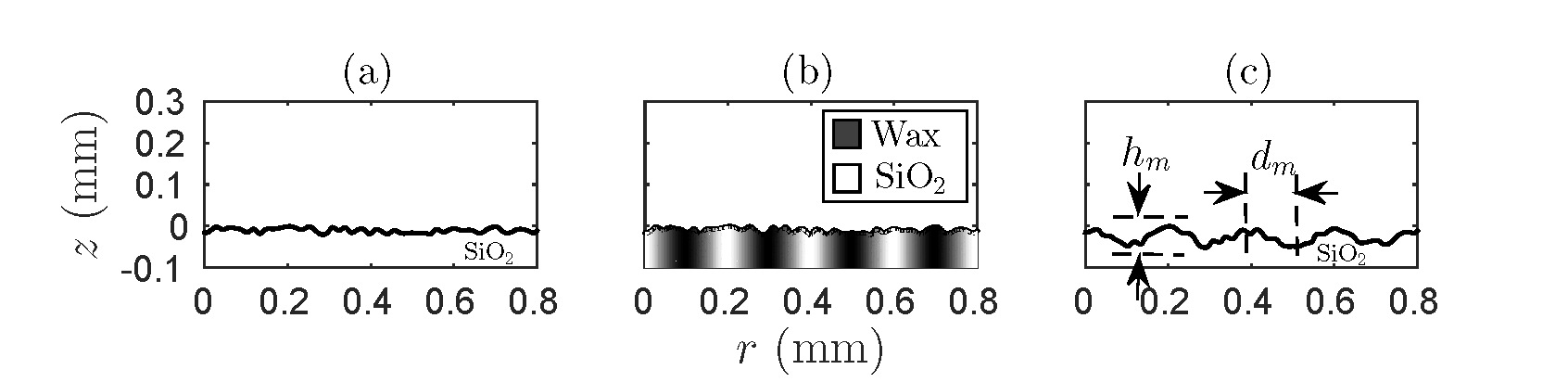}
\caption{Solid substrates featuring: (a) micro-scale textures ($r_f$ = 1.15),  (b) micro-scale textures accompanied with chemical heterogeneities and (c) hierarchical roughness with micro- and macro-scale textures.}
\label{fig9}
\end{figure}

As experimentally observed\cite{biance2004, Bird2008} for low viscosity fluids on smooth and perfectly wetting substrates, there exists an inertial regime where the contact radius grows with time according to the power law:
\begin{equation}
\frac{R_s} {R_0} = K \left( \frac{t} {t_c} \right) ^ {1/2}, \label{power_law}
\end{equation}
where $K$ is the power law prefactor and $t_c$ is the inertio-capillary time ($t_c=\sqrt{\frac{\rho R_0^3}{\gamma}}$). The above applies for $t/t_c<1$, where the dynamics are inertially dominated. Stapelbroek et al.\cite{stapelbroek2014} experimentally demonstrated that the geometrical morphology and the chemical heterogeneities of the substrate play a negligible role during the early spreading case. In particular, even if the final equilibrium profiles are different, they argue that there exist an ubiquitous inertial regime where the power law (Eq.~(\ref{power_law})) is applicable. Here, in order to demostrate this inertial spreading universality, we perform computations of the previously examined glycerin/water mixture droplet spreading on different types of solid surfaces. Indicatively, we examine the effect of geometrical in conjunction with with chemical heterogeneities on silicon dioxide (SiO{$_\text{2}$}) substrates (see Fig.~(\ref{fig9})). In the first solid substrate case (see Fig.~(\ref{fig9}a)), the micro-scale (intrinsic) geometrical complexity corresponds to a roughness factor of 1.15, which equals to the minimum roughness required to match the experimental measurements in the case of spreading on a horizontal wax surface (see Fig.~(\ref{fig3})). 
\begin{figure}
\includegraphics[scale=0.7]{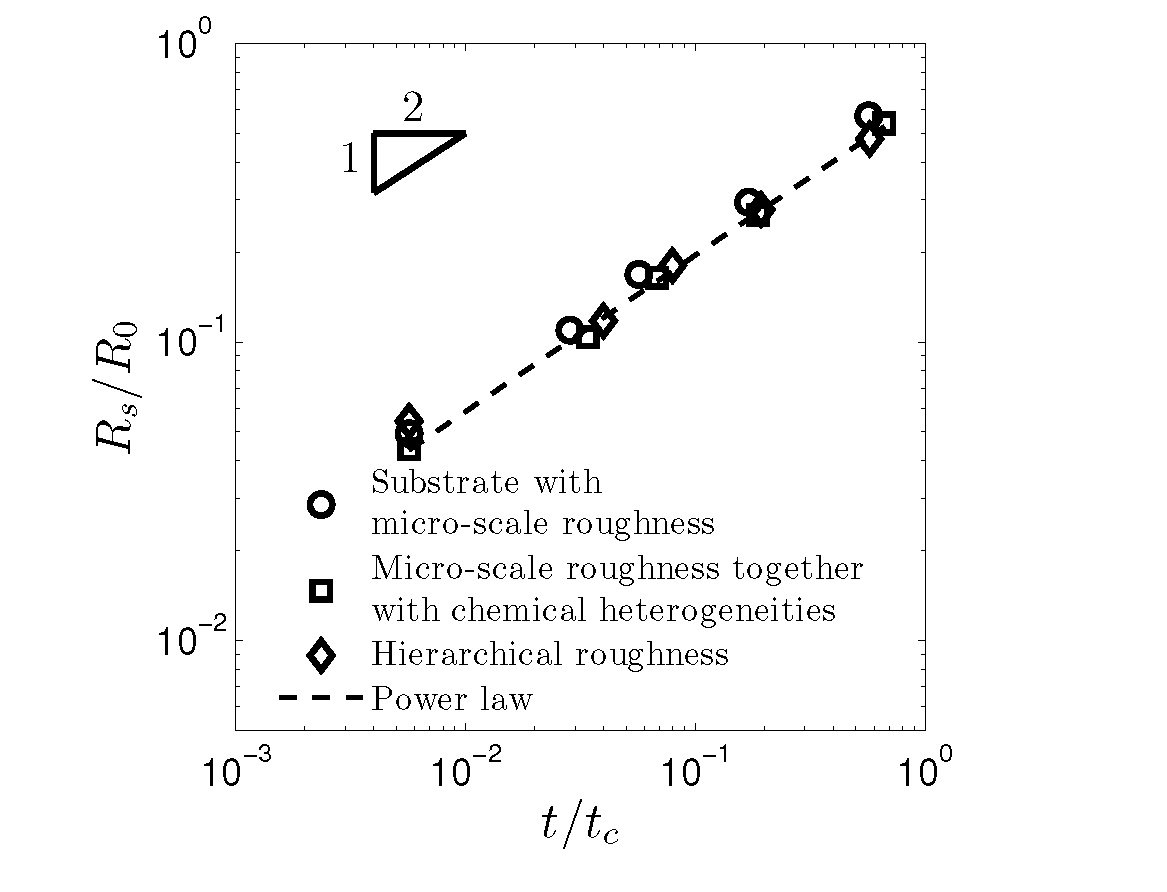}
\caption{Normalized contact radius, $R_s/R_0$, as a function of dimensionless time from impact, $t/t_c$, for three different types of solid substrates featuring topographical and chemical inhomogeneities (see Fig.~(\ref{fig9})).}
\label{fig10}
\end{figure}
Chemical patterns are introduced in the second solid surface case, by alternating the hydrophilic SiO{$_\text{2}$} substrate with wax patches (see Fig.~(\ref{fig9}b)). Indicatively, the wetting parameter, $w^{LS}$, varies along the spatial $r$ coordinate, ranging from a minimum (SiO{$_\text{2}$}) to a maximum value (wax) with a wavelength of 0.2 mm. Note that the wettability of the SiO{$_\text{2}$} corresponds to a $\theta_Y$ = 5\textsuperscript{o}, and $\theta_Y$ = 93.5\textsuperscript{o} for the wax substrate. Lastly, a hierarchical roughened solid surface (see Fig.~(\ref{fig9}c)), where the micro-scale intrinsic roughness is superimposed on larger sinusoidal structures (macro-scale roughness), is also investigated. Regarding the hierarchical patterned surface (Fig.~(\ref{fig9}c)), the first to second level roughness amplitude ratio is $\frac{h_m}{h}=1.7$,  whereas the distance of the respective inhomogeneities correspond to a ratio $\frac{d_m}{d}=6.8$ ($h_m$ = 0.033 mm and $d_m$ = 0.1 mm). The impact velocity of the droplet is now neglected since the approach speed at the relative experiments is extremely low ($u_0< 20 \times 10^{-6}$ m/s) \cite{stapelbroek2014}. 
\begin{figure}
\includegraphics[scale=0.7]{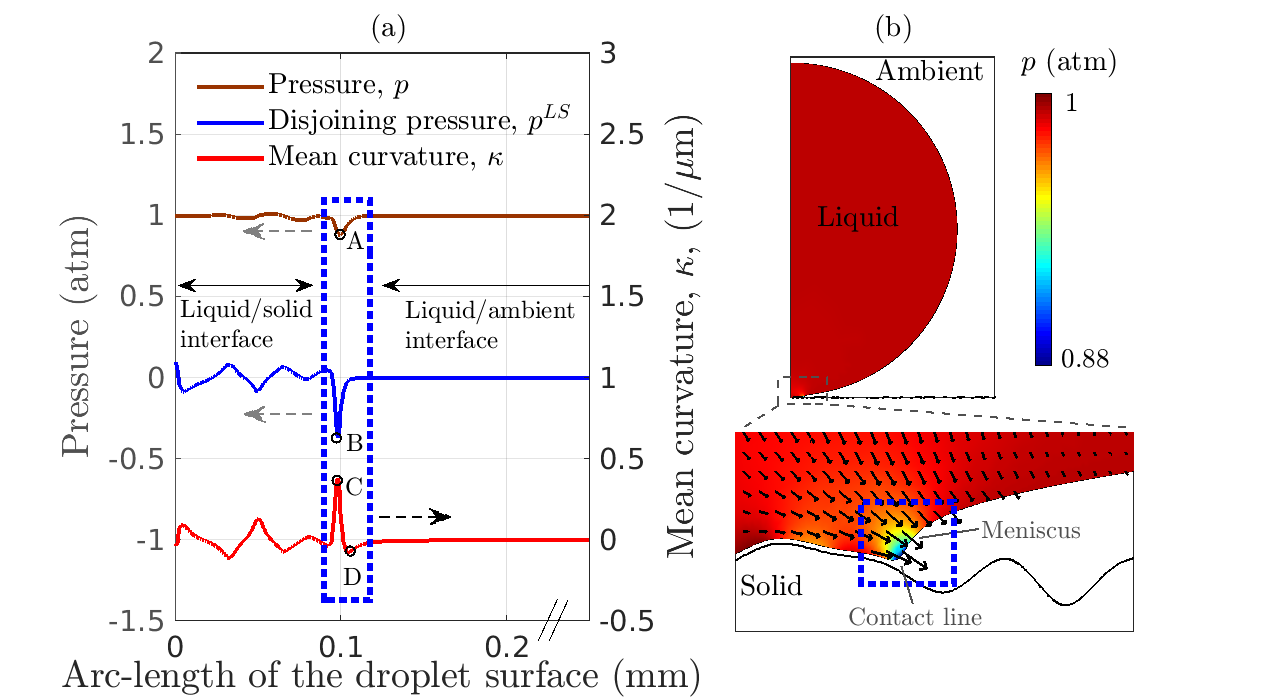}
\caption{(a) Mean curvature, $\kappa$, (right axis) and pressure (left axis) values along the effectively one-dimensional droplet surface just after the contact ($t$ = 5 $\times 10^{-2}$ ms) with a rough ($r_f$ = 1.15) SiO{$_\text{2}$} substrate. (b) Visualization of the pressure, $p$, distribution for the entire droplet and a magnified region close to the contact line. The arrows at the latter sub-figure represent the fluid velocity field, $\boldsymbol{u}$.}
\label{fig11}
\end{figure}
\par
As demonstrated in Fig.~(\ref{fig10}), the evolution of the normalized contact radius, $R_s/R_0$ is indeed identical for the different types of surface complexity during the early stage of spreading $(t/t_c<1)$. The results are plotted in logarithmic scale so as to demonstrate the the power law (Eq.~(\ref{power_law})) growth dynamics of the early spreading inertial regime.
In order to interpret the negligible role of the underlying substrate to the spreading dynamics, we revisit the normal component of Eq.~(\ref{navierbcondition}) which reads: 
\begin{equation}
\boldsymbol{n} \cdot \boldsymbol{\tau} \cdot \boldsymbol{n} = - p_{ext} -2 \gamma \kappa , {\text{at SQ.}} \label{stress_balance_normal}
\end{equation}
By neglecting the deviatoric stress component of $\boldsymbol{\tau}$, due to the negligible impact velocity, the stress balance along the droplet surface (Eq.~(\ref{stress_balance_normal})) for the early spreading stage can be approximated by:
\begin{equation}
-p \approx - p_{ext} -2 \gamma \kappa \xRightarrow{\text{Eq.~(\ref{externalpressure})}} p - p_0 \approx p^{LS} + 2 \gamma \kappa, {\text{at SQ,}} \label{curvature-pressure}
\end{equation}
where $p_0$ = 1 atm.
When the droplet equilibrates, the pressure difference between the liquid and the ambient medium, $p - p_0$, is maintained constant along the effectively one-dimensional droplet surface. In our case, however, the calculation of the right hand side of Eq.~(\ref{curvature-pressure}), $p^{LS} + 2 \gamma \kappa$, along the entire droplet profile at the time instance just after impact ($t$ = 5 $\times 10^{-2}$ ms), shows a significant deviation from a constant value, localized in the vicinity of the solid surface. This results to a local pressure drop, close to the solid substrate region, as illustrated in Fig.~(\ref{fig11}). Namely, the highly negative disjoining pressure value at the contact line (equals to the depth of the Lennard-Jones potential well; see Eq.~(\ref{disjoiningpressure})) (point B in Fig.~(\ref{fig11}a)) is not compensated by the surface tension forces, $2 \gamma \kappa$, despite the fact that $\kappa$ reaches a peak value (point C in Fig.~(\ref{fig11}a)). This unbalanced surface force generates a pressure gradient inside the droplet (see point A in Fig.~(\ref{fig11}a)), which drives the initial liquid motion over topographical or chemical inhomogeneities, annihilating the role of the substrate. At equilibrium, the pressure difference between the liquid and the surrounding medium, $p - p_0$, should approach zero since the extremely small Young contact angle ($\theta_Y$ = 5\textsuperscript{o}) drives the droplet into a liquid film. This yields to: $p^{LS} = - 2 \gamma \kappa$, at SQ, from Eq.~(\ref{curvature-pressure}). Previous theoretical attempts to explain the spreading insensitivity on the roughness structure\cite{wu2004, stapelbroek2014} suggest that the strong curvature induced at the liquid meniscus, connecting the droplet and the substrate (see Fig.~(\ref{fig11}b)), is the origin of the pressure difference. However, in our detailed analysis, by examining the entire droplet surface, we conclude that the curvature of the meniscus does not have a major contribution; contrariwise the key factor is the stress balance at the contact line.
Specifically, we calculate that just after the droplet contacts the substrate ($t$ = 5 $\times 10^{-2}$ ms), the absolute curvature value of the liquid meniscus equals to $7.8 \times 10^{-2}$ $\mu$m\textsuperscript{-1} (point D in Fig.~(\ref{fig11}a)) whereas the corresponding value at the contact line is $ 38 \times 10^{-2}$ $\mu$m\textsuperscript{-1} (point C in Fig.~(\ref{fig11}a)). To our knowledge, this is the first time that this complex interfacial phenomenon is successfully captured by a modified hydrodynamic modeling approach.
%
%
\section*{Summary and conclusions}
Conventional hydrodynamic models require the implementation of an explicit slip boundary condition at the contact line (e.g. the correlations of Cox\cite{Cox1986}, Hoffman-Voinov-Tanner law\cite{hoffman1975, voinov1976, tanner1979}) to avoid the nonphysical stress singularity. Therefore, when simulating a droplet impact on a geometrically structured solid surface, a contact angle boundary condition has to be applied to multiple and unknown in number contact lines - an essentially tedious or even infeasible task. In order to tackle these severe limitations, we propose a novel modeling approach which treats the liquid/solid, as well as the liquid/ambient interfaces in a unified manner. Contrary to the conventional models, in our computations the dynamic contact angle, as well as the liquid/solid relative motion, are derived implicitly as a result of the complex flow occurring in the vicinity of the solid surface intrinsic roughness, governed by micro-scale liquid/solid interactions. The above are internally introduced in the Navier-Stokes equations with a disjoining pressure term, rendering the boundary condition implementation at the contact line, albeit necessary at the conventional hydrodynamic models, redundant in our approach.
\par
Aiming to validate our methodology, we examined the dynamic behavior of an axisymmetric droplet spreading on a horizontal surface. We found that when enhancing the micro-scale intrinsic roughness of the solid surface, our results gradually converge to the experimental measurements. Interestingly enough we demonstrated that local viscous forces, which are generated at the solid surface roughness length scale, give rise to effective (macroscopic) shear stresses resisting the droplet deformation. The above is accompanied with a substantial energy loss at the contact line in the case of a roughened substrate, contrary to an ideally smooth solid surfaces, as noticed by visualizing the local rate of energy dissipation.
\par
By studying the initial spreading behavior of a droplet on highly composite - hierarchically and/or chemically patterned - solid surfaces, we concluded that there exists an inertial regime where the contact radius evolution is independent of the underlying solid substrate complexity. During this time interval, the spreading dynamics can be well approximated by a universal power law  (Eq.~(\ref{power_law})). This argument is in remarkable agreement with experimental observations performed by Stapelbroek et al. in a recently published work\cite{stapelbroek2014}. At later times, the only factor that limits the dynamics is the viscosity of the liquid; this slow spreading regime can also be captured by our model, in agreement with Tanner's law\cite{tanner1979}.
\par
Regarding the future perspectives, our computations enable the derivation of an overall energy dissipation term (including all the micro-scale dissipation phenomena taking place on the solid roughness) as a function of the contact line velocity. Such a phenomenological approach, which is a subject of ongoing research, could greatly simplify the computations since the complex roughness geometry will be replaced by a flat boundary featuring the same friction properties; different roughness scales can also be simulated in this way. These kind of approximations are already used in modified hydrodynamic models\cite{ren2007} as well as in macro-scale phase-field computations\cite{carlson2012, wang2015}, where the microscopic interactions at the contact line are incorporated through a single friction parameter. The proposed modeling approach is however advantageous over these methodologies when dealing with complex surface topographies.
\par
With nature as starting point (systems with self-cleaning properties such as lotus leafs\cite{nosonovsky2009}, highly adhesive rose petals\cite{feng2008} or even desert lizards with the ability to transport water over their skin\cite{sherbrooke2007}), many researchers have focused on studying the wetting behavior of droplets on roughened substrates and then designing structure geometries to obtain desirable wetting properties\cite{cavalli2012parametric, pashos2015modified}. Active control of the wetting dynamics (e.g. the acceleration of the spreading process by modifying the electric potential of the droplet\cite{courbin2009, chen2013} or by imposing a temperature gradient on the substrate\cite{karapetsas2014}) has recently also gained attention as it is related with a lot of technical applications (e.g. inkjet printing, lab-on-a-chip devices, etc.).
In this direction, and since the incorporation of additional forces (e.g. the electric field effect, Marangoni stresses) on the surface stress balance can be readily performed in our approach, it could be interesting to study the control of spreading dynamics by combining an external force action and the solid surface geometric characteristics effect. Future work also focuses on extending our model to account for the surrounding medium flow, by solving a separate set of Navier-Stokes equations, in order to capture more complex impact phenomena e.g. the development of a thin air layer below the droplet\cite{mandre2012, de2015}.
%
%
\begin{acknowledgments}
The authors kindly acknowledge funding from the European Research Council under the Europeans Community's Seventh Framework Programme (FP7/2007-2013)/ERC grant agreement no. [240710]. We are also grateful to Dr. A. Petsi from FORTH/ICE-HT and Dr. G. Karapetsas from the National Technical University of Athens for the useful discussions on the subject.
\end{acknowledgments}
\begin{appendix}
%
%
\section{Eikonal equation}
\begin{figure}
\includegraphics[scale=0.7]{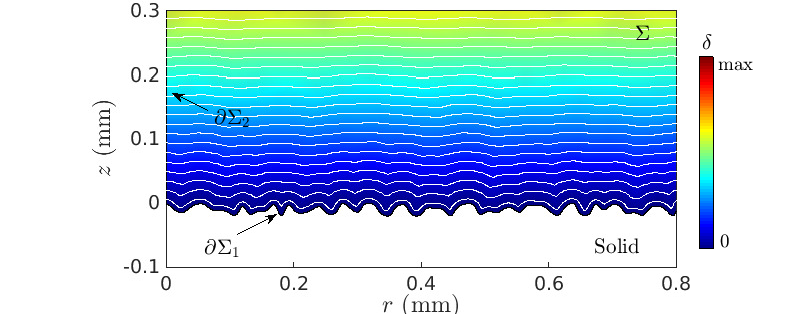}
\caption{Solution of the Eikonal equation and contour lines of constant distance, $\delta$, for an arbitrary roughened solid surface. Note that for the better viewing of the roughness details, only a part of the domain is presented (the upper and the right boundaries are located a $z$ = 3 mm and $r$ = 3 mm respectively, in order to cover the entire droplet).}
\label{fig12}
\end{figure}
The disjoining pressure, $p^{LS}$, is a function of the Euclidean distance from the solid boundary. For geometrically structured solid surfaces (see Fig.~(\ref{fig2})), this distance is obtained from the solution of the Eikonal equation, which reads in two-dimensional form\cite{dacorogna1999}:
\begin{subequations}
\label{eikonal}
\begin{eqnarray}
\mid \nabla \delta (r,z) \mid = 1,\quad r,z \in \Sigma, \label{eikonal_equation}
\\
\delta = 0, \quad r,z \in \partial \Sigma_1, \label{eikonal_equation_bc1}
\\
\nabla \delta  \cdot \boldsymbol{ n_s}= 0, \quad r,z \in \partial \Sigma_2, \partial \Sigma_3, \partial \Sigma_4, \label{eikonal_equation_bc2}
\end{eqnarray}
\end{subequations}
where $\Sigma$ denotes the computational domain; $\partial \Sigma_1$ is the solid surface boundary and $\partial \Sigma_2$, $\partial \Sigma_3$, $\partial \Sigma_4$ are the left, the upper and the right boundaries, respectively (see Fig.~(\ref{fig12})). The vector $\boldsymbol{n_s}$ is the unit normal to the boundary. Among the proposed solution schemes we adopt the approach of Fares and Schr{\"o}der\cite{fares2002}. Specifically Eq.~(\ref{eikonal_equation}) is modified to:
\begin{equation}
\nabla D \cdot \nabla D + \beta D (\nabla \cdot \nabla D) = (1+2\beta)D^4, \label{eikonal_modified}
\end{equation}
where $D = \frac{1}{\delta}$ and $\beta$ is a numerical diffusion parameter, which enhances the elliptic behavior of the equation. The solution of the above equation is more accurate as $\beta$ approaches zero, however extremely low values can hinder its convergence. In our computations $\beta = 5 \times 10^{-3}$. Since the inverse distance, $D$, reaches infinity at the solid boundary, the respective boundary condition is\cite{fares2002}:
\begin{equation}
D = \frac{C}{l_{ref}}, \quad r,z \in \partial \Sigma_1, \label{eikonal_modified_bc}
\end{equation}
where the constant $C =2$ and $l_{ref}$ depends on the geometry of the computational domain (equals to the half of the shortest side of the domain). The Eikonal equation (Eq.~(\ref{eikonal_modified})) is discretized and solved in COMSOL Multiphysics\textsuperscript{\textregistered} only once for each solid surface geometry, therefore the added computational cost to the overall problem is negligible.
%
%
\section{Effect of the disjoining pressure parameters}
\begin{figure}
\includegraphics[scale=0.7]{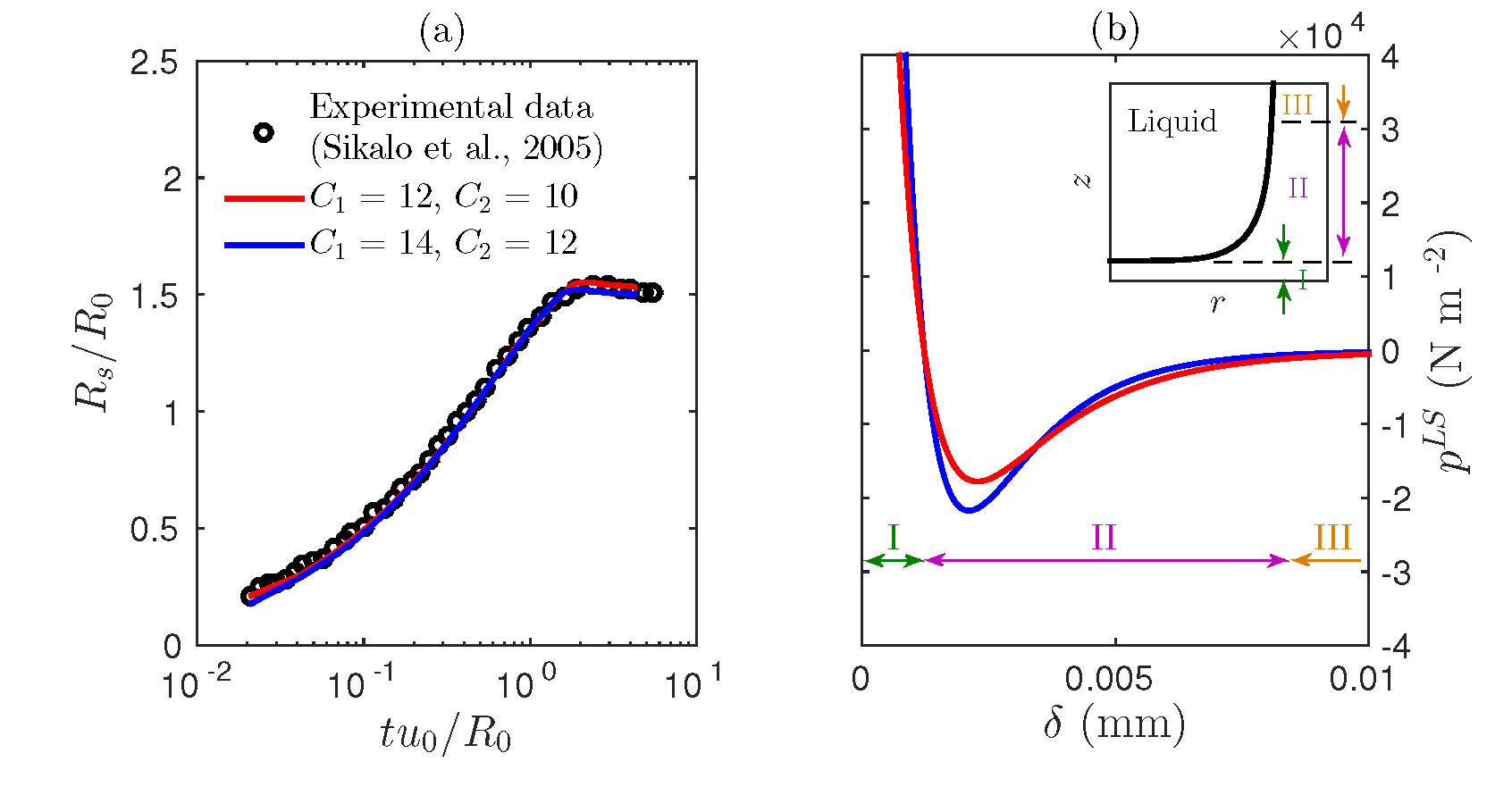}
\caption{(a) Normalized contact radius of a glycerin/water mixture droplet impacting on wax surface: simulations against experimental data from Sikalo et al.\cite{sikalo2005} for different $C_1$ and $C_2$ parameters of the disjoining pressure (see Eq.~(\ref{disjoiningpressure})). The arising disjoining pressure profiles are illustrated in (b) as a function of the distance from the solid boundary, $\delta$. The inset in (b) shows the action range of the micro-scale forces in the vicinity of the contact line; attractive interactions are dominant in region (II), whereas repulsive forces are generated in region (I), keeping the liquid and solid phases separated. For $\delta \approx$ 0.01 mm and beyond (region (III)), the micro-scale liquid/solid interactions are negligible.}
\label{fig13}
\end{figure}
\begin{figure}
\includegraphics[scale=0.7]{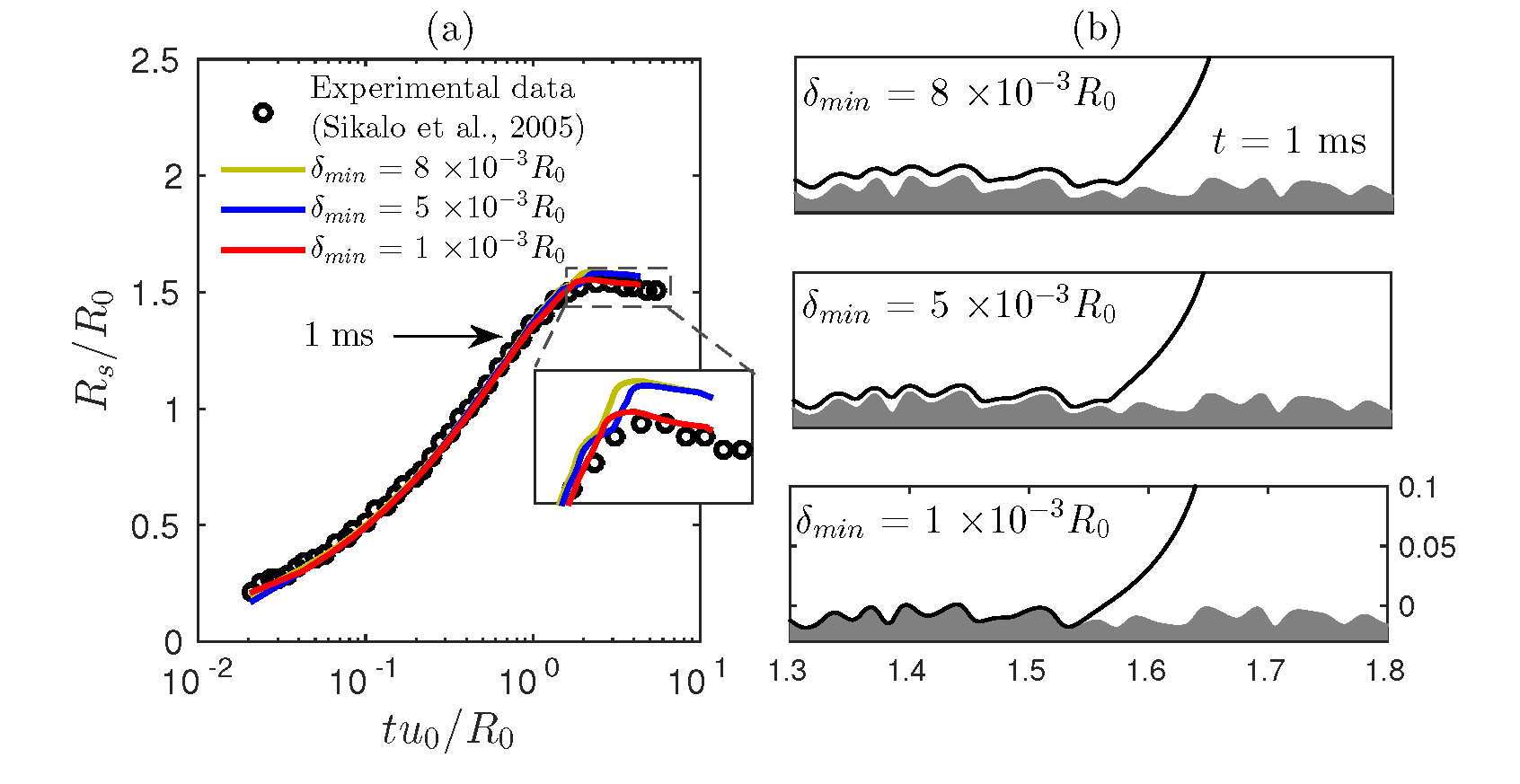}
\caption{(a) Normalized contact radius evolution of a glycerin/water mixture droplet impacting on wax surface: simulations against experimental data from Sikalo et al.\cite{sikalo2005} for different minimum distances ($\delta_{min} \in [1 \times 10^{-3} R_0, 8 \times 10^{-3} R_0]$) between the solid and the liquid phase. The corresponding droplet profiles for the different values of $\delta_{min}$ are demonstrated in (b) at $t$ = 1 ms.}
\label{fig14}
\end{figure}
In this section, we investigate whether a modification in the parameters $C_1, C_2$ and $\sigma$ of the disjoining pressure (see Eq.~(\ref{disjoiningpressure})) has an impact on the spreading behavior of a droplet on a solid substrate. In all cases we study a glycerin/water mixture droplet spreading on a wax solid surface with a roughness factor, $r_f$ = 1.15. In particular, in Fig.~(\ref{fig13}) by comparing the contact radius evolution for two different $C_1, C_2$ parameters sets ($C_1 = 12, C_2 =10$ and $C_1 = 14, C_2 =12$), we conclude that the results are in remarkable agreement despite the different disjoining pressure profiles (an increase in the exponents $C_1, C_2$ leads to a narrower range of micro-scale interaction as shown in Fig.~(\ref{fig13}b)). Next, in  Fig.~(\ref{fig14}) we investigate the effect of the minimum distance, $\delta_{min}$, between the liquid and solid phases, which encompasses the impact of $\sigma$ and $\epsilon$ as follows: $\delta_{min} = R_0 (\sigma - \epsilon)$. Specifically, starting from a large initial value ($\delta_{min} = 8 \times 10^{-3} R_0  = 9.8 \times 10^{-3}$ mm) we reduce the minimum distance by lowering the parameter $\sigma$ ($\sigma \in [9 \times 10^{-3}, 16 \times 10^{-3}]$) keeping constant $\epsilon =  8 \times 10^{-3}$. 
Conclusively, as presented in Fig.~(\ref{fig14}a), the computational results converge to the experimental data by reducing $\delta_{min}$ to its minimum value ($\delta_{min} = 1 \times 10^{-3} R_0 = 1.225 \times 10^{-3}$ mm), which is obtained for $\sigma = 9 \times 10^{-3}$.
%
%
\section{Mesh deformation and refinement}
\begin{figure}
\includegraphics[scale=0.7]{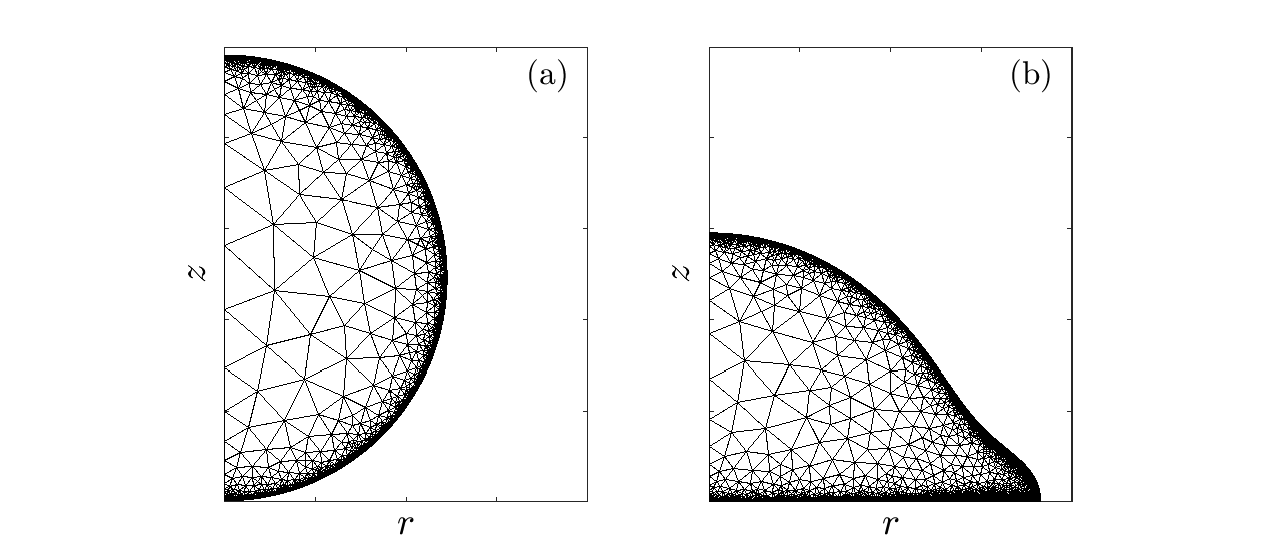}
\caption{(a) Initial and (b) deformed computational mesh of an axisymmetric droplet impacting on a perfectly smooth solid surface.}
\label{fig15}
\end{figure}
\begin{figure}
\includegraphics[scale=0.7]{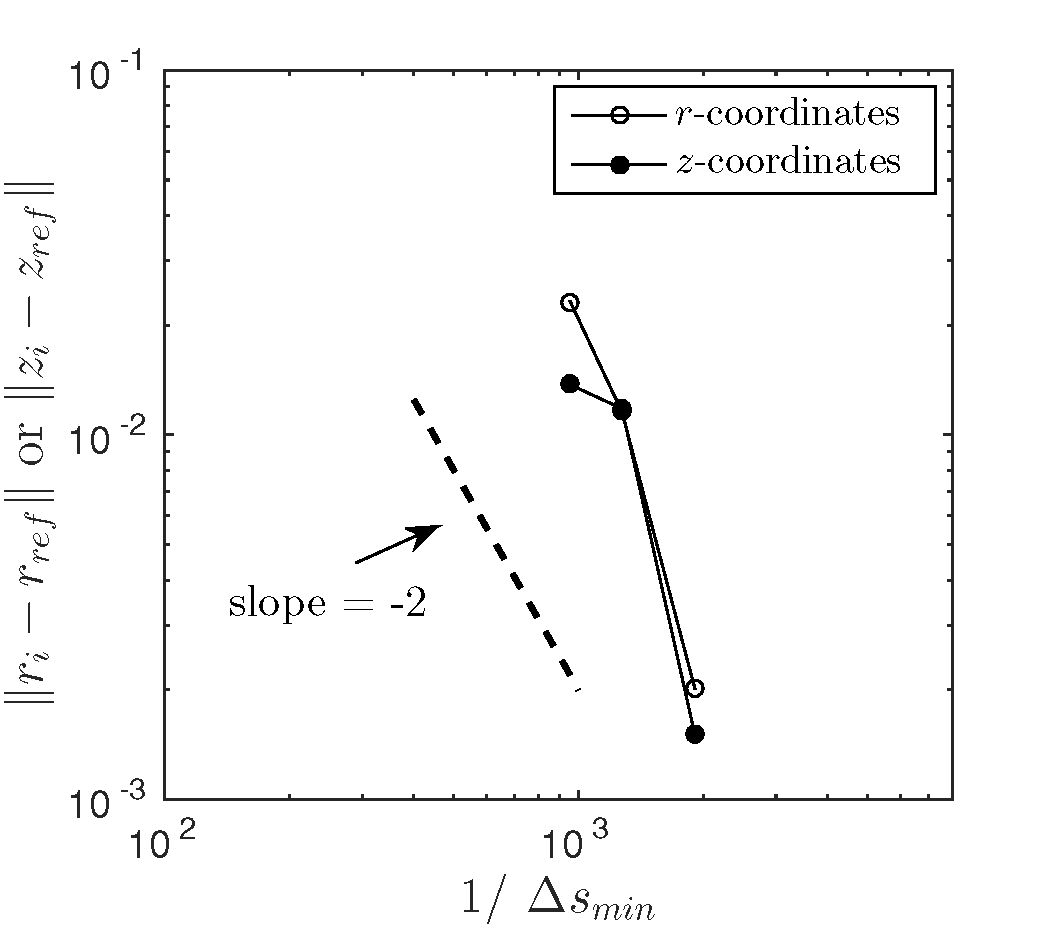}
\caption{The relative error of the droplet shape coordinates, when impacting on a rough wax surface ($r_f$ = 1.2, $t$ = 5 ms) for various meshes. A reference solution with $7 \times 10^3$ elements along the droplet surface is used (the length of the smallest computational element, $\Delta s_{min}$ = $0.45 \times 10^{-3}$ mm).}
\label{fig16}
\end{figure}
In order to adapt the  two-dimensional unstructured mesh to the deformed droplet geometry we solve the Winslow smoothing equations\cite{Knupp1999} for the spatial mesh points coordinates $r$ and $z$ of the droplet interior:
\begin{subequations}
\label{winslow_equations}
\begin{eqnarray}
\frac{\partial	^2 \xi}{\partial r^2} + \frac{\partial ^2 \xi}{\partial z^2} =0, \label{winslow_equations_r}
\\
\frac{\partial	^2 \eta}{\partial r^2} + \frac{\partial ^2 \eta}{\partial z^2} =0, \label{winslow_equations_z}
\end{eqnarray}
\end{subequations}
accounting for the kinematic boundary condition (Eq. (\ref{meshvelocity})), where $\xi$ and $\eta$ refer to a material coordinate system following the deformation of the domain.
Material coordinates ($\xi$, $\eta$) initially coincide with the spatial ($r$, $z$) until the droplet surface is deformed. In Fig.~(\ref{fig15}) we demonstrate the initial and a deformed computational mesh of a droplet impacting on a horizontal solid surface.
\par
The location of the droplet interface is tracked by the deformed mesh and thus high mesh density is required along the boundary of the domain. The accuracy of the results is tested against discretization refinement for the case of a glycerin/water mixture droplet impacting on the most roughened substrate ($r_f$ = 1.2). In particular in Fig.~(\ref{fig16}) we depict the relative error of the droplet shape coordinates $(r, z)$ for different meshes, compared with a reference solution. The reference solution has the denser mesh with $7 \times 10^3$ computational elements along the droplet surface (note that a typical size of the computational problem is of the order of $10^5$ degrees of freedom, including the velocity field, pressure and mesh deformation). For the calculation of the norms, the values of the droplet shape coordinates $(r, z)$ were interpolated at 250 points which are placed on equal distances throughout the droplet surface of the coarsest mesh. Next, the $r$, $z$ coordinates were interpolated for the other meshes, at the locations corresponding to the 250 arc-length coordinates of the coarsest mesh. Finally, as presented in Fig.~(\ref{fig16}) the convergence with mesh refinement is super-quadratic.
%
%
\section{Viscous spreading: Validation with Tanner's law}
\begin{figure}
\includegraphics[scale=0.7]{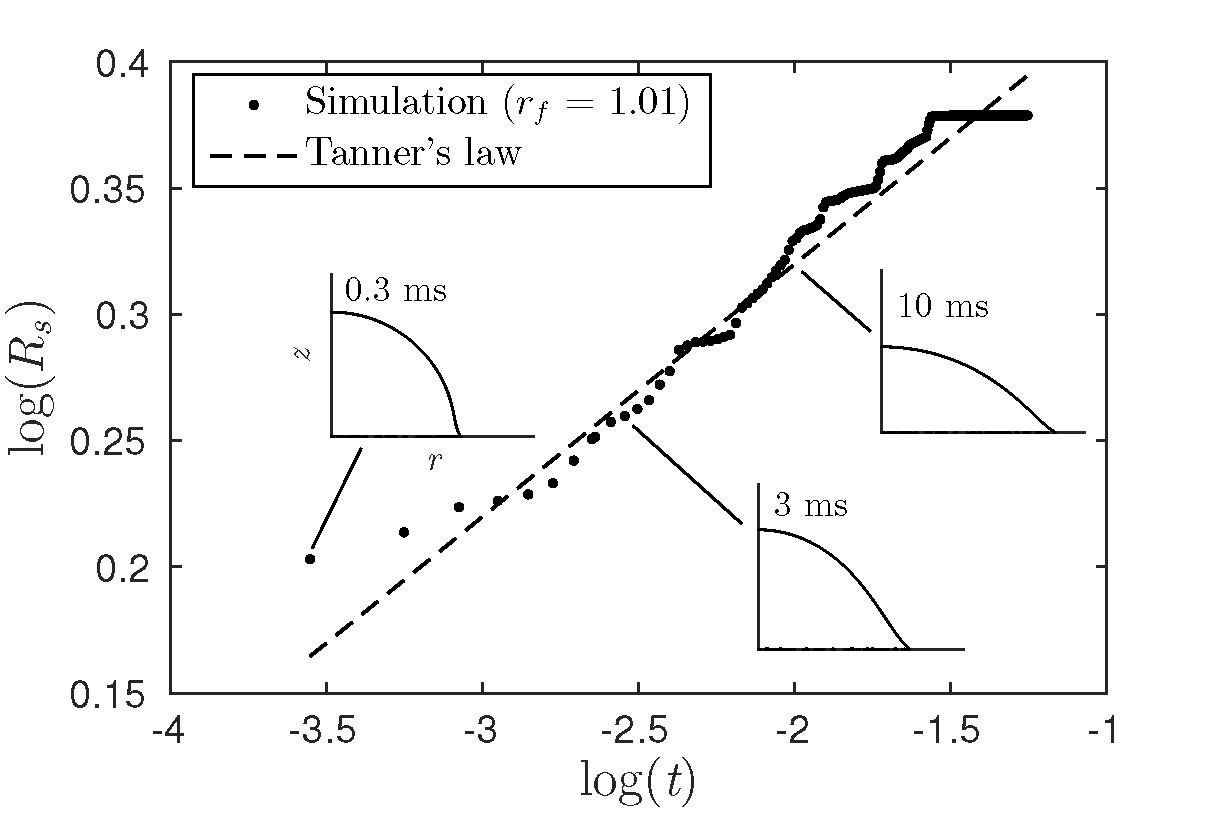}
\caption{Evolution of the contact radius as a function of time for a glycerin/water mixture droplet on a hydrophilic substrate ($\theta_Y$ = 30\textsuperscript{o}).}
\label{fig17}
\end{figure}
In the case of wettability driven spreading, the radius of the wetted area can be described by the well-known Tanner's law\cite{tanner1979}:
\begin{equation}
R_s \sim t^{1/10}, \label{Tanner}
\end{equation}
where the only factor that limits spreading is the viscous dissipation near the contact line. Eq.~(\ref{Tanner}) has been experimentally verified\cite{biance2004} and is an excellent problem to benchmark our modeling approach for viscous-dominated dynamics. In Fig.~(\ref{fig17}) we demonstrate the dynamic behavior of a glycerin/water mixture droplet spreading on a hydrophilic substrate ($\theta_Y$ = 30\textsuperscript{o}), assuming an intrinsic surface roughness of $r_f$ = 1.01. Starting from an equilibrium solution of $\theta_Y$ = 93.5\textsuperscript{o}, it is observed that our simulations can successfully capture Tanner's law at intermediate times ($t \in (1, 26)$ ms). For $t>$ 26 ms the droplet reaches equilibrium and is finally immobilized.
\end{appendix}
%
%
\newpage
\bibliography{ref}
%
%
\end{document}